\documentclass[10pt,aps,prl,twocolumn,groupedaddress,showkeys]{revtex4-2}
\usepackage{amsmath}
\usepackage{graphicx}
\usepackage[colorlinks=true,linkcolor=blue,urlcolor=blue,citecolor=blue,anchorcolor=blue]{hyperref}
\DeclareGraphicsExtensions{.pdf,.eps,.png,.jpg,.mps}
\usepackage{epstopdf}
\usepackage{threeparttable}
\usepackage{color}
\usepackage[T2A]{fontenc}
\usepackage[utf8]{inputenc}
\usepackage{float}
\usepackage{tabularx}
\usepackage{array}
\usepackage[font=small,labelfont=bf]{caption}
\usepackage[percent]{overpic}

\begin{document}

\title{Broadband, Flexible, Skin-Compatible Carbon Dots/Graphene Photodetectors for Wearable Applications}
\author{Nouha Loudhaief$^1$}
\author{Petr Rozhin$^1$}
\author{Ilaria Bertuol$^1$}
\author{Ali Raza$^{1,2}$}
\author{Leonardo Viti$^3$}
\author{Subhankar Roy$^3$}
\author{Enrico Cavarzerani$^1$}
\author{Luca Sbuelz$^2$}
\author{Matteo Brilli$^1$}
\author{Andrea Serinolli$^1$}
\author{Jacopo Nicoletti$^1$}
\author{Riccardo Piccoli$^1$}
\author{Sebastián Castilla$^4$}
\author{Simone Dal Zilio$^2$}
\author{Miriam Serena Vitiello$^3$}
\author{Maurizio Selva$^1$}
\author{Flavio Rizzolio$^{1,5}$}
\author{Alvise Perosa$^1$}
\author{Giovanni Antonio Salvatore$^1$}
\author{Domenico De Fazio$^1$}
\email[]{domenico.defazio@unive.it}
\affiliation{$^1$ Department of Molecular Science and Nanosystems, Ca’ Foscari University of Venice, Via Torino 155, 30172 Venice, Italy}
\affiliation{$^2$ CNR-IOM Istituto Officina dei Materiali, S.S. 14-Km. 163, 5–34149 Trieste, Italy}
\affiliation{$^3$ NEST, CNR-Istituto Nanoscienze and Scuola Normale Superiore, Piazza San Silvestro 12, 56127 Pisa, Italy}
\affiliation{$^4$ ICFO-Institut de Ciències Fotòniques, The Barcelona Institute of Science and Technology, Avinguda Carl Friedrich Gauss 3, 08860 Castelldefels (Barcelona), Spain}
\affiliation{$^5$ Pathology Unit, Centro di Riferimento Oncologico di Aviano (C.R.O.) IRCCS, Via Franco Gallini 2, 33081 Aviano, Italy}

\keywords{photodetectors, carbon dots, graphene, flexible devices, wearable}

\begin{abstract}
The development of wearable photodetectors demands a unique combination of broadband optical sensitivity, mechanical flexibility, and skin-compatibility, with these requirements rarely met simultaneously by existing technologies. Here, we present photodetectors that combine all of these performances. This is achieved by integrating carbon dots, engineered for extended ultraviolet-to-near-infrared absorption, with single-layer graphene transferred onto a plastic substrate. Unlike traditional quantum dot systems, our carbon dots achieve a broad ultraviolet-to-near-infrared response without toxic heavy metals. Graphene provides an efficient channel for charge transport, while a biocompatible chitosan-glycerol electrolyte enables efficient, low-voltage carrier modulation, with peak performance at approximately 0.5 V gate bias. The resulting photodetectors exhibit a broadband photoresponse with responsivities of approximately 0.19 A/W at 406 nm, 0.32 A/W at 642 nm, and 0.18 A/W at 785 nm. They maintain consistent performance at a bending radius of 0.8 cm with negligible degradation after repeated cycles. Furthermore, skin-compatibility assessments using the SkinEthic\textsuperscript{\textnormal{TM}} model confirm the non-toxic nature and suitability of our devices for direct skin contact. The combination of broadband absorption (400-800 nm), flexibility, and skin-compatibility, along with low-voltage operation ($<$ 1.5 V), positions our photodetectors as promising building blocks for next-generation wearable optoelectronics. 
\end{abstract}

\maketitle

\section{\label{Intro}Introduction}
Photodetectors (PDs) are fundamental building blocks in optoelectronics, converting incident light into electrical signals. In the $\sim$200-1100 nm wavelength range, conventional PDs, primarily based on crystalline silicon (Si) \citep{ReedWiley2008, JalaliJLightwaveTechnol2006, ThomsonJOpt2016}, have long served as the backbone of optoelectronic systems due to their technological maturity, high performance, and cost-effectiveness, and are seamlessly integrated onto complementary metal–oxide–semiconductor (CMOS) platforms where mechanical flexibility is not required \citep{ReedWiley2008, JalaliJLightwaveTechnol2006, ThomsonJOpt2016}. However, the accelerating demand for flexible and wearable electronic and photonic components for applications such as real-time biometric monitoring, alongside the rapid expansion of Internet of Things (IoT) systems, is exposing the limitations of rigid PD technologies \citep{WeinpjFlexElectron2024, ShajariSensors2023, YanNanoMicroLett2025}. To address the transition towards flexible devices, a new class of PDs is emerging that combines broadband operation (ultraviolet-to-near-infrared sensitivity) with mechanical flexibility and biocompatibility, all essential features for conformal and safe integration with the human skin and natural environments \citep{LinSensdiagn2024, LombarteFrontBiotechnol2021, ZhuBiomaterials2023}. These features are critical for advancing next-generation platforms in wearable healthcare \citep{PolatSciAdv2019, LochnerNatCommun2014, LouAdvSci2023, KumarACSSens2025}, smart textiles \citep{ChoiAdvEngMater2025}, environmental sensing \citep{KozlowskiProCompSci2023}, and distributed IoT nodes \citep{ZaiaAdvElectMater2019, XiaoAdvMeas2020}.

For health-monitoring related applications, broadband spectral response is particularly important for multispectral sensing tasks. Photoplethysmography (PPG) \citep{ChenSensors2020} and pulse oximetry \citep{KimInfDisp2021}, for example, rely on visible (VIS) red and near-infrared (NIR) light for accurate cardiovascular monitoring, while ultraviolet (UV) detection supports skin health assessment and environmental exposure monitoring \citep{KaragiorgisNpjFlexElectron2025, HorshamJMIR2020}. Beyond these requirements, next-generation PDs must also maintain high optoelectronic performance, characterized by key metrics. For example, responsivity $R$, measured in A/W or V/W, is a key figure of merit that quantifies the amount of electrical signal generated per unit of incident optical power \citep{ZouPhotonics2024}. Alongside $R$, the response time, typically assessed via the rise and fall times ($\tau_\text{r}$ and $\tau_\text{f}$, respectively), provides insight into the detector’s temporal resolution and switching dynamics \citep{WangNatCommun2023}. Achieving a balance between good optoelectronic performance and application-driven requirements, such as flexibility, broadband operation, skin-compatibility, and potential scalability, would be key to enabling reliable, efficient wearable systems for applications such as continuous monitoring and personalized diagnostics \citep{LiuSciRep2024, CaiAdvMater2019}.

While numerous flexible PDs have been proposed in recent years, many focus on optimizing only one or a few of these key figures of merit, which can limit their suitability for wearable and skin-integrated applications \citep{LiuSciRep2024, WangMaterSciEngRRep2023}. For example, zinc oxide (ZnO)-based PDs are known for their exceptional $R$ up to $1.1 \times 10^{6}$ A/W under UV illumination (365 nm), but they remain spectrally narrowband and typically exhibit slow response times ($\tau_\text{r} \sim 10$ s, $\tau_\text{f} \sim 12$ s), which limits their utility for wearable real-time sensing \citep{KaragiorgisNpjFlexElectron2025}. Organic photodetectors extend their response into the UV-to-NIR (400-950 nm) region with $R$ peaking at $\sim$ 0.47 A/W at 800 nm in Ref. \citenum{SchrickxAdvOptMater2024} and offer mechanical flexibility, but they lack verified long-term biocompatibility, often require complex multi-layer fabrication, and inevitably experience performance degradation over time from material instability \citep{KaushalBiosensors2023}. Perovskite photodetectors offer a potentially broadband response with peaking $R$ $\sim$ 0.15-0.2 A/W at 610-650 nm in Ref. \citenum{AzamatACSApplOptMater2024}, but are prone to degradation when exposed to heat and moisture, and they contain highly toxic components. Quantum dots (QDs) are known for their strong light absorption, low-cost fabrication, and size-tunable bandgap \citep{ZhanLightSciAppl2025}, but as isolated units, they suffer from poor charge carrier mobility due to charge hopping-dominated transport \citep{JiangJMaterChem2022}, which limits their ability to extract photogenerated carriers efficiently. Most importantly, they are typically composed of toxic elements such as lead (Pb) or cadmium (Cd) \citep{Reach, DengACSApplMaterInterfaces2025}. 

In this scenario, two-dimensional (2D) materials are widely emerging as a class of materials that offer complementary advantages, including wide spectral response, mechanical flexibility, gate-tunability, high carrier mobility, and more, making them attractive for optoelectronic applications \citep{JiangJMaterChem2022, NovoselovScience2004, BonaccorsoNatPhotonics2010, KonstantatosNatCommun2018}. However, being formed by one or few atomic layers, their light absorption is intrinsically limited, necessitating integration with other light-absorbing materials for efficient photodetection \citep{WangMaterResExp2019, CaoMicromachines2021, FanSciRep2018, NairScience2008}. This drawback is often overcome by integrating graphene into hybrid PD architectures with other low-dimensional (0D, 1D, or 2D) materials, thereby preserving its excellent charge transport properties while enhancing light absorption \citep{AbbasMicrosystNanoeng2024}. One outstanding solution relies on graphene/QD hybrids \citep{PolatSciAdv2019, ZhangAdvOpt2022, ChanLowDimSysNano2020, LiACSApplMaterInterfaces2025}, yet the presence of heavy metals in their composition (Pb, Cd) raises concerns for skin contact and environmentally sustainable wearable applications \citep{Reach, TrifirograveMathewsJPharmSci2024, MiaoFrontEnergyRes2021}. For example, Ref. \citenum{LiACSApplMaterInterfaces2025} combined graphene-QD devices with eco-friendly substrates to obtain very high $R$ $\sim$8$\times 10^{4}$ A/W at 520 nm and millisecond switching times ($\tau_\text{r} \sim 51$ ms, $\tau_\text{f} \sim 338$ ms), but QDs in the heterostructure contained Pb. Collectively, these examples highlight the lack of a PD that simultaneously meets the multiple requirements needed for broadband, flexible, and skin-integrable applications.

To address this challenge, we developed a broadband, flexible, skin-compatible, potentially scalable, and gate-tunable hybrid PD by integrating low-cost, scalable, hydrothermally synthesized carbon dots (CDs) with chemical vapor deposition (CVD)-grown single layer graphene (SLG) on a polyethylene terephthalate (PET) substrate, gated via a biopolymer-based chitosan–glycerol (CS-GL) electrolyte. CDs are emerging 0D nanomaterials known for their strong intrinsic UV absorption, solution processability, and excellent biocompatibility \citep{OliveiraMaterLet2021, ZuoMicrochimActa2016, TaspikaRSCAdv2019}. Unlike toxic QDs, CDs are carbon-based, environmentally benign, and their optical response can potentially be tuned via surface functionalization to extend absorption into VIS and NIR regions \citep{MagdySciRep2023, YuLightAdvManuf2024}. To enable broadband response, the CDs in this work were thus optimized to extend their absorption into the NIR region, thereby achieving a photoresponse in the $\sim$400-800 nm range. Meanwhile, the CS-GL gate electrolyte, a flexible and biocompatible biopolymer \citep{AhmadACSApplMater2025}, was chosen for its ionic conductivity, enabling stable electric double-layer (EDL) gating while maintaining skin safety \citep{SharovaNanoscale2023}. Moreover, both CDs and chitosan used here can be derived from biowaste through upcycling processes \citep{OliveiraMaterLet2021, AhmadACSApplMater2025}. This synergistic architecture leverages CDs as broadband light absorbers covering the UV-VIS-NIR range, SLG as the high-mobility transport channel, and the soft-gel CS-GL electrolyte for low-voltage ($<$ 1.5 V) EDL gating to modulate the properties of our PD, all integrated on a lightweight PET substrate widely used in wearable electronics \citep{NanMicromachines2022, HassanMicromachines2025, HassanJMaterSci2025}. The resulting PD exhibits a stable UV-VIS-NIR photoresponse with $R$ of approximately 0.2, 0.3 and 0.2 A/W at 406, 642, and 785 nm, respectively, peaking at a gate bias of around 0.5 V. Further, the PD demonstrates $\tau_\text{r}$ of 0.3-1.1 s and $\tau_\text{f}$ of 1.7-2.1 s, which are well within the temporal requirements for wearable and on-skin sensing applications \citep{SharovaLabChip2018, ChenACSSens2020}. Finally, SkinEthic\textsuperscript{\textnormal{TM}} assays confirm the non-toxic nature of our PD, supporting safe on-skin integration. The combination of broadband optical response, mechanical adaptability, skin-safe operation, and low-voltage functionality within a single platform paves the way for truly wearable optoelectronic systems.

\section{\label{Results}Results and Discussion}

\subsection{\label{Opto}Optoelectronic Performance}

\begin{figure*}[htbp!]
\centerline{\includegraphics[width=180mm]{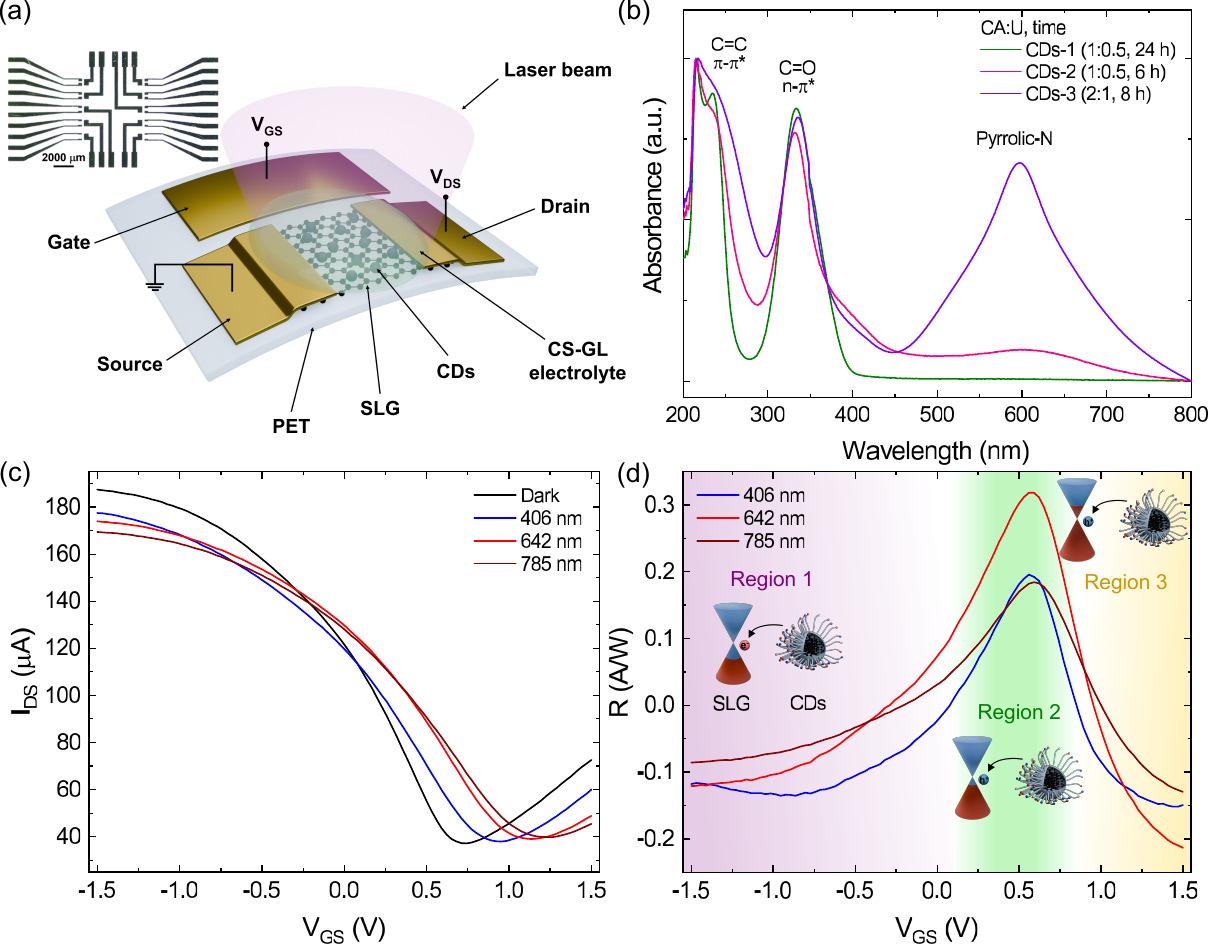}}
\caption{(a) Schematic illustration of the CDs/SLG/PET photodetector gated with CS-GL electrolyte; inset: optical image of a sample containing 10 devices. (b) UV-VIS absorption spectra of CDs synthesized under different conditions for NIR optimization, varying the mass ratio of citric acid (CA) to urea (U) and the reaction time. (c) Transfer characteristics of the device in dark and illuminated conditions at different wavelengths, measured at $V_\mathrm{DS}$ = 100 mV. (d) Gate-dependent responsivity extracted from the transfer curves in (c), with Region 1 (violet), Region 2 (green), and Region 3 (yellow). Device area: 150 $\mu$m × 320 $\mu$m.}
\label{fig:Figure1}
\end{figure*}

Figure \ref{fig:Figure1}(a) shows the device schematics. An array of ten devices with channel lengths of 100-300 $\mu$m and a width of 320 $\mu$m was fabricated, as shown in the optical microscopy image inset of Figure \ref{fig:Figure1}(a). Each device consists of a SLG channel clamped between the source and drain electrodes. A gate electrode is integrated next to each device to enable individual gating. A solvent-resistant grade of PET was selected as the flexible substrate due to its properties, including transparency of $\sim$90\% throughout the visible spectrum, chemical resistance to solvents commonly employed in microfabrication processes (e.g., acetone and isopropyl alcohol), thermal stability, and mechanical robustness \citep{FarajSciAdv2017, BaeNatNanotechnol2010}. Commercially available CVD-SLG (Graphenea), grown on copper foil, was transferred onto PET using a poly(methyl methacrylate) (PMMA)-assisted wet transfer method, as detailed in the Experimental Section \citep{BaeNatNanotechnol2010}. The quality of the SLG and how it is affected by the transfer procedure was assessed by Raman spectroscopy, at an excitation wavelength of 514 nm, on the as-grown film on copper foil and after transfer onto a SiO$_2$/Si substrate, to avoid confusion arising from the Raman-active modes of PET, that may hinder those of graphene (see Supplementary Figure \ref{fig:FigureS1}) \citep{LiACSNano2015}. The gold source, drain, and gate electrodes were patterned on the SLG/PET using a shadow mask (see Experimental Section). 

CDs, the main photoactive component in the device, were synthesized via a hydrothermal method using citric acid as the carbon source and urea as nitrogen dopant to enhance absorption in the NIR region. We synthesized an extensive series of CD samples by systematically varying the synthesis parameters, including precursor ratio, temperature, and reaction time. Detailed synthesis conditions are provided in the Experimental Section. Among these, Figure \ref{fig:Figure1}(b) shows three representative samples to illustrate the evolution of their absorption spectra. CDs-1 (olive curve), prepared with a citric acid:urea mass ratio of 1:0.5 and reacted for 24 h, show negligible absorption beyond 400 nm. CDs-2 (pink curve), obtained with the same 1:0.5 mass ratio but a shorter reaction time of 6 h, exhibits the emergence of a new absorption band centered around 600 nm, with an intensity $\sim$16-fold higher than CDs‑1. Finally, CDs-3 (violet curve), synthesized with a higher citric acid:urea mass ratio of 1:2 and an 8 h reaction time, displays a pronounced absorption peak at 600 nm, with a $\sim$7-fold enhancement compared to CDs-2 ($\sim$109-fold compared to CDs-1) and a tail largely extending into the NIR. Specifically, the absorption spectrum of CDs-3 exhibits three characteristic bands \citep{TaspikaRSCAdv2019, PermatasariACSApplNanoMater2018}: (i) a high-energy peak at approximately 240 nm, ascribed to $\pi$–$\pi$* electronic transitions within aromatic C=C bonds located in the sp$^{2}$-hybridized conjugated domains of the carbon core; (ii) a mid-UV band near 335 nm, attributed to n–$\pi$* transitions originating from surface carbonyl (C=O) groups; (iii) a broad absorption feature, centered around 600 nm, extending into the NIR window, which we ascribe to electronic transitions involving surface-localized states rich in pyrrolic nitrogen \citep{TaspikaRSCAdv2019, PermatasariACSApplNanoMater2018}. We then integrated CDs-3 samples into our PD architecture due to their UV-to-NIR ($\sim$400-800 nm) absorption, preserving biocompatibility and environmental safety. A droplet of aqueous CD dispersion (1:0.1 w/v, CDs:ultrapure water) was cast onto the graphene channel using a micropipette and allowed to dry on a hot plate at 60 °C to remove residual moisture. Additional characterization of the CDs, including Raman spectroscopy, transmission electron microscopy (TEM), and Fourier-transform infrared spectroscopy (FTIR), was performed to gain insight into their structural, morphological, and chemical features. These analyses confirmed the presence of disordered carbon, nanoscale particle size ($\sim$7 nm), and surface functional groups on the CDs (see Supplementary Figure \ref{fig:FigureS2} and \ref{fig:FigureS3}).

Finally, a biocompatible chitosan-glycerol (CS-GL) electrolyte, whose preparation is described in the Experimental Section, was selected \citep{AhmadACSApplMater2025}, drop-cast over the channel region, and left to dry overnight under ambient conditions to form the top-gate layer. Chitosan, a non-toxic biopolymer derived from chitin, offers excellent film-forming ability, aqueous processability, and environmental sustainability \citep{FengSciRep2016, MinSciRep2020}. Its abundant hydroxyl and amine groups facilitate proton conduction, allowing the formation of high-capacitance electric double layers (EDLs) at the graphene interface \citep{FengSciRep2016, MinSciRep2020}. The incorporation of glycerol into the chitosan matrix, as a biocompatible plasticizer, further enhances the ionic conductivity and flexibility of the electrolyte by disrupting the chitosan crystalline structure and increasing ion mobility \citep{AsnawiPolymers2020}. In contrast to conventional polymer electrolytes used with graphene, such as polyethylene oxide-lithium chlorate (PEO-LiClO$_{4}$)\citep{KinderACSAppl2017}, which rely on toxic lithium and perchlorate ions, the CS-GL system provides a safer, metal-free alternative, well suited for integration into biocompatible optoelectronic platforms. This study presents the first demonstration of a CS-GL electrolyte used to gate SLG for photodetection applications. 

To further evaluate the optical contribution of each layer, UV-VIS spectra were recorded after the sequential deposition of the individual materials. As shown in Supplementary Figure \ref{fig:FigureS4}, the introduction of CDs significantly enhances the absorption across the VIS-NIR regions. Specifically, the absorbance around 350 nm increases by more than 3-fold compared to the SLG/PET, while around 600 nm, it exhibits a $\sim$10-fold enhancement. These results confirm that the CDs are the dominant photo-absorbing component in our devices. Additionally, to validate the role of SLG and CDs in broadband photodetection, we performed control output and photoresponse measurements on SLG/PET and CDs/PET devices (See Supplementary Figure \ref{fig:FigureS5}). Taken together, these measurements show that CDs alone are non-conductive, while SLG, although electrically conductive, exhibits only a minimal intrinsic photoresponse. This highlights the necessity of a hybrid configuration, where CDs act as efficient light-harvesting components that transfer photogenerated carriers to graphene, acting as a conducting channel for photodetection.

Figure \ref{fig:Figure1}(c) shows the transfer characteristics of the device under dark conditions and illumination at 406, 642, and 785 nm, measured at $V_\mathrm{DS}$ = 100 mV. The $\pm$1.5 V gate range reflects the electrochemical stability window of the CS-GL electrolyte, wide enough to ensure efficient channel modulation via the EDL, yet narrow enough to prevent electrolyte degradation \citep{FengSciRep2016, HamsanPolymers2021}. In the dark, the representative device exhibits ambipolar behavior with the charge neutrality (Dirac) point located at $\sim$0.75 V, while across multiple devices the Dirac point was typically found at 0.93 $\pm$0.18 V, indicating slight device-to-device variations and confirming the intrinsic p-type doping of the ungated SLG channel, which can arise from several commonly reported factors in graphene-based devices, including the adsorption of atmospheric species such as oxygen and water molecules \citep{Panchal2DMater2016}, substrate-induced charge transfer \citep{PiazzaCarbon2016}, and residual contaminants introduced during fabrication \citep{LupinaACSNano2015}. Another critical factor is the choice of electrode material, as the difference in the work function between graphene and the contact metal can modulate the Fermi level of the channel \citep{ParkACSNano2015}. In our devices, gold electrodes were used. Due to its relatively high work function ($\sim$5.1 eV) \citep{MichaelsonACSNano2015}, gold tends to extract electrons from graphene, thus inducing stronger p-type doping. In contrast, when silver paint was applied manually in a test configuration (see Supplementary Figure \ref{fig:FigureS6}), the Dirac point for the representative device shown in that figure was observed around 0.05 V, indicating a significantly lower level of p-doping, consistent with the lower work function of silver($\sim$4.3-4.7 eV) \citep{MichaelsonACSNano2015}.

Under illumination, the transfer characteristics reveal a complex photoresponse resulting from charge-transfer processes between the CDs and the SLG channel. This behavior can be divided into three distinct regions, defined by the crossing points between the dark and illuminated curves. For clarity, we define these regions based on the gate-voltage ranges extracted from the transfer characteristics under each illumination wavelength. Region 1 is defined approximately as -1.5 V < $V_\mathrm{GS}$ < 0.07 V for 406 nm, -1.5 V < $V_\mathrm{GS}$ < -0.29 V for 642 nm, and -1.5 V < $V_\mathrm{GS}$ < -0.2 V for 785 nm. Region 2 extends from 0.07 V to 0.86 V (406 nm), -0.29 V to 0.95 V (642 nm), and -0.2 V to 1.0 V (785 nm). Region 3 spans from the upper bounds of Region 2 up to 1.5 V. These regions are visually distinguished in Figure \ref{fig:Figure1}(d) by violet, green, and yellow background shading, corresponding to Regions 1, 2, and 3, respectively, which displays the responsivity as a function of the gate voltage for the three excitation wavelengths (406, 642, and 785 nm). The responsivity $R$ is calculated using the relation: $R = I_\mathrm{ph}$/$P_\mathrm{eff}$, where \( I_\mathrm{ph} \) is the photocurrent, and \( P_\mathrm{eff} \) is the effective optical power incident on the device, calculated as \citep{ZouPhotonics2024}: $P_\mathrm{eff} = P_\mathrm{laser}A_\mathrm{device}/A_\mathrm{spot}$, where $P_\mathrm{laser}$ is the total output power of the laser, $A_\mathrm{device}$ is the active area of the device, and $A_\mathrm{spot}$ is the beam spot area, larger than $A_\mathrm{device}$. These responsivity curves were derived from the transfer characteristics shown in Figure \ref{fig:Figure1}(c), by calculating the photocurrent at each gate voltage as the difference between the illuminated and dark current. The light was focused using a plano-convex lens, and $A_\mathrm{spot}$ was measured using a blade \citep{AraujoApplOpt2009}; the beam radius was found to be approximately 0.78 mm for all wavelengths. To further clarify the underlying mechanisms, schematic diagrams illustrating the charge transfer processes between the CDs and the SLG channel in the three distinct gate-voltage regions are also included in Figure \ref{fig:Figure1} (d).

In Region 1, the device operates in the p-type conduction regime, where holes are the majority carriers. Upon illumination, CDs absorb photons and generate electron-hole pairs. The photoexcited electrons are transferred from the CDs to the SLG, where they partially recombine with the hole population, effectively reducing the number of holes available for conduction. As a result, the channel becomes more resistive, and the current decreases. In Region 2, the device remains p-type but exhibits reversed behavior compared to Region 1. Here, illumination leads to an increase in current, implying that photoexcited holes from the CDs are transferred to the SLG and actively contribute to conduction. This additional hole injection enhances the p-type conductivity of the channel. The corresponding positive shift of the Dirac point under illumination supports this interpretation, as it indicates a downward shift of the Fermi level due to the increased hole population in SLG. As highlighted in Figure \ref{fig:Figure1}(d), $R$ reaches its maximum in this region, with values of approximately 0.19 A/W (406 nm), 0.32 A/W (642 nm), and 0.18 A/W (785 nm), all occurring near $V_\mathrm{GS} = 0.5$ V. We infer that this peak in responsivity arises from an optimal balance between dark current near the Dirac point and efficient photogenerated hole injection from the CDs. In Region 3, the device transitions to the n-type conduction regime, where electrons dominate transport. Under illumination, photoexcited holes from the CDs are still transferred to the SLG, where they recombine with electrons, thereby reducing the net carrier density and consequently decreasing the current.

To estimate the specific detectivity $(D^*)$, another key metric in PDs, we employed the following expression \citep{FuAdvDevicesInstrum2023}:
\begin{equation}
    D^* = R{\frac{\sqrt{A_\mathrm{device}}}{i_\mathrm{noise}}}
\end{equation}
where $R$ is the responsivity, and $i_\mathrm{noise}$ is the noise current, measured in units of $\mathrm{A/\sqrt{Hz}}$, and estimated as the square root of the dark current power spectral density $S_\mathrm{I}$ at 100 mHz (see Supplementary Figure \ref{fig:FigureS7}). The resulting $(D^*)$ are approximately $3.6 \times 10^5$ Jones (406 nm), $6.0 \times 10^5$ Jones (642 nm), and $3.4 \times 10^5$ Jones (785 nm). These relatively low values of $D^*$ may be attributed to the $\sim\mu{A}$ dark current in our device. This issue may be resolved in future work by replacing SLG with a semiconducting channel, while trying not to compromise the highly mobile charge transport.

It is worth mentioning that, given the amorphous structure of CDs, as confirmed by TEM (Supplementary Figure \ref{fig:FigureS3}(a)), and their defect-rich nature, a conventional band alignment model may not fully capture the photocarrier dynamics at the interface with SLG. However, based on the photoresponse trend observed in Figure \ref{fig:Figure1}(c) and Figure \ref{fig:Figure1}(d), we may infer that in Region 1, the energetic landscape favors electron transfer from the photoexcited CDs to the SLG. As the gate voltage increases and the Fermi level shifts upward, electron injection becomes less favorable, and hole transfer instead becomes dominant, as observed in Regions 2 and 3.

\begin{figure*}[htbp!]
\centerline{\includegraphics[width=180mm]{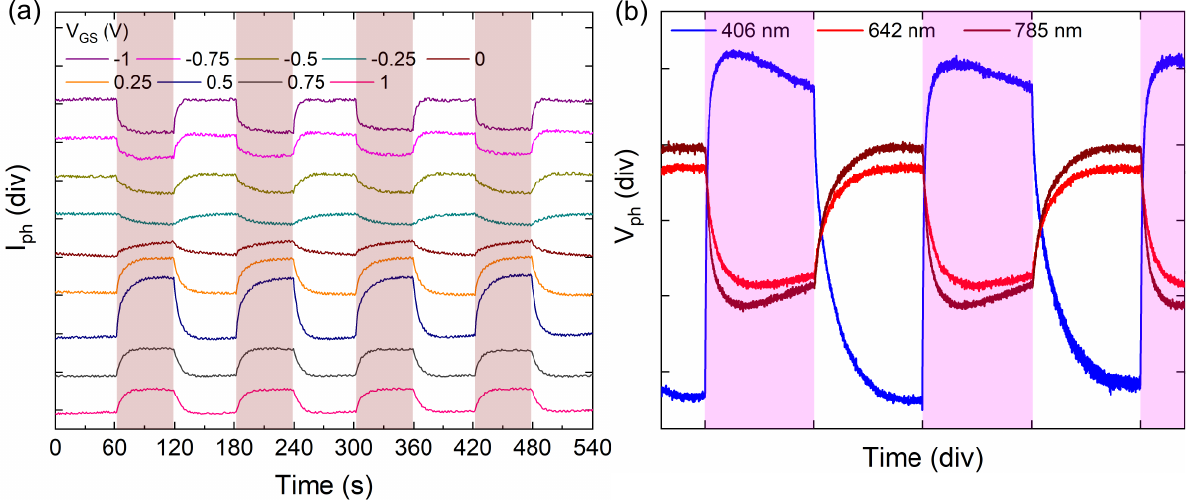}}
\caption{(a) Time-dependent photoresponse under different gate voltages at 785 nm, measured with a fixed $V_\mathrm{DS}$ = 50 mV. The curves are vertically shifted for clarity. The vertical scale is 2 $\mu$A/div. Shaded dark red regions indicate the laser ON states. Device area: 150 $\mu$m × 320 $\mu$m. (b) Oscilloscope-recorded photovoltage ($V_\mathrm{ph}$) response under 100 mHz modulated illumination at various wavelengths, measured at zero gate bias under constant $I_\mathrm{DS} = 50\ \mu$A. The horizontal scale is 5 s/div, and the vertical scale is 20 mV/div. Magenta regions correspond to laser ON states. Device area: 250 $\mu$m × 320 $\mu$m.}
\label{fig:Figure2}
\end{figure*}

As shown in Figure \ref{fig:Figure1}(c), illumination at different wavelengths leads to a positive shift of the Dirac point; however, the extent of this shift varies with wavelength, reaching 0.95, 1.13, and 1.25 V for 406, 642, and 785 nm, respectively. We attribute this variation primarily to differences in ${P_\mathrm{eff}}$. Specifically, ${P_\mathrm{eff}}$ on the active device area was 0.09 mW for 406 nm, 0.11 mW for 642 nm, and 0.21 mW for 785 nm, chosen based on the available laser output and to ensure stable operation (see Experimental Section for details on the laser source). Higher incident power enhances the generation of photoexcited carriers within the CDs, thereby increasing the rate of hole transfer to the SLG and resulting in a stronger photodoping effect \citep{ChengSciRep2013}. This photodoping effect, which refers to the light-induced modulation of carrier density in graphene, leads to hole accumulation, lowering the Fermi level and shifting the Dirac point toward more positive gate voltages \citep{LinSciRep2016, TetsukaSciRep2017}. To quantify this photodoping effect, we estimate the photo-induced carrier density ($\Delta n$) using the expression $\Delta n = (C_\mathrm{tot} \Delta V_\mathrm{Dirac})/q$, where $C_\mathrm{tot}$ is the total capacitance per unit area between the gate and the graphene channel, {$\Delta V_\mathrm{Dirac}$} is the shift of Dirac point in the transfer characteristics under illumination compared to dark conditions, and $q$ is the elementary charge \citep{DeFazioACSNano2016}. To determine $C_\mathrm{tot}$, electrical impedance spectroscopy (EIS) was performed using a two-terminal configuration, where the drain and source electrodes of the device were electrically shorted and connected to ground. A small sinusoidal AC voltage with an amplitude of 10 mV was applied to the gate electrode, and the resulting current response was recorded over a frequency range from 1 MHz down to 1 kHz. The measured impedance spectrum was then fitted using an equivalent R-(RC) circuit model (Figure \ref{fig:FigureS8}), yielding $C_\text{tot} = 9.58 \times 10^{-8}~\mathrm{F{\cdot}cm^{-2}}$. Using this capacitance and the observed Dirac point shift under different wavelength illumination, we estimate $\Delta n$ values of $5.68\times 10^{11}$, $6.76 \times 10^{11}$, and $7.48 \times 10^{11}\,\mathrm{cm}^{-2}$ under 406, 642, and 785 nm illumination, respectively. Beyond quantifying the amount of $\Delta n$, these values are also crucial for estimating the photoconductive gain, $G$, of the device, which represents the number of charge carriers circulating through the external circuit per single absorbed photon \citep{YangCoatings2023}. In photogated graphene PDs, light absorbed by the sensitizer generates charges that remain trapped with a lifetime significantly exceeding the carrier transit time in graphene, thereby sustaining electrostatic doping of the graphene channel and modulating its conductivity. A single trapped photogenerated charge can therefore allow many carriers to flow through the graphene before it is neutralized. The gain can be expressed in terms of the measured photocurrent as \citep{DeFazioACSNano2016}:
\begin{equation}
    G = \frac{I_\mathrm{ph}}{q A_\mathrm{device} \Delta n}
\end{equation}

At a gate voltage of 0.5 V, the calculated $G$ are \(3.7\times10^{5}\), \(6.8\times10^{5}\), and \(6.7\times10^{5}\) at 406, 642, and 785 nm, respectively. The obtained $\Delta n$ and corresponding $G$ are consistent with other graphene-hybrid photodetectors exploiting photogating. For instance, Ref. \citenum{DeFazioACSNano2016} reported polymer-electrolyte-gated SLG/MoS$_2$ PDs on PET with similar $G$ values on the order of 10$^{5}$. Slightly lower $G$ values, on the order of 10$^{4}$, were reported for Bi$_2$Te$_3$ nanowire/graphene hybrids in Ref. \citenum{YooNanomaterials2021}. Conversely, higher $G$ values, on the order of 10$^{6}$-10$^{8}$, were estimated for perovskite nanocrystal/graphene PDs in Ref.  \citenum{CottamACSApplElectronMater2020} and lead sulfide (PbS) QDs/graphene phototransistors in Ref. \citenum{KonstantatosNatNanotechnol2012}, suggesting longer carrier trapping times in the perovskite nanocrystals and QDs, respectively, compared to CDs.

\begin{figure*}[htbp!]
\centerline{\includegraphics[width=180mm]{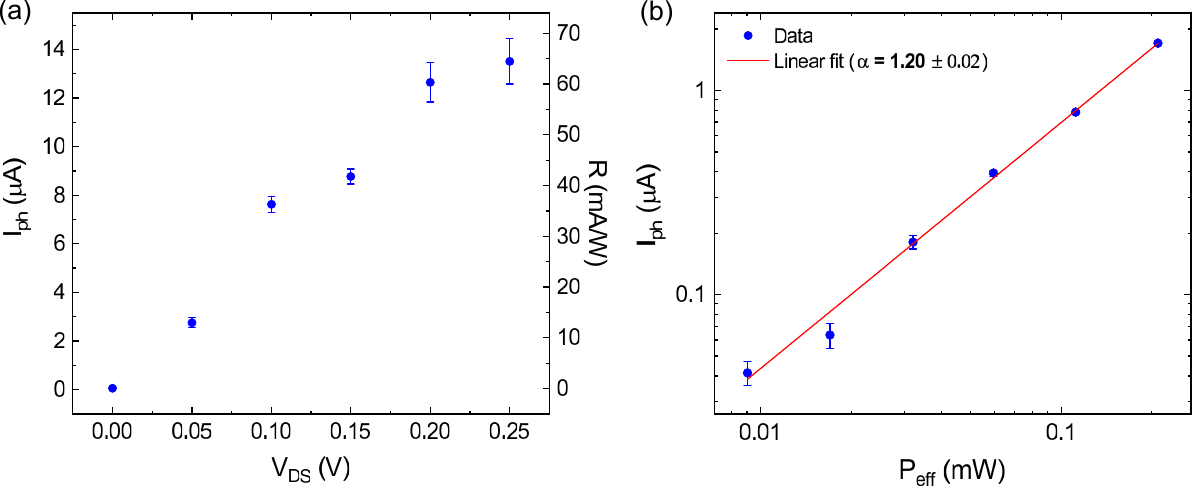}}
\caption{(a) Source-drain voltage-dependent photocurrent (left axis) and corresponding responsivity (right axis). (b) Effective optical power (${P_\mathrm{eff}}$)-dependent photocurrent (${I_\mathrm{ph}}$). Measurements were performed at 785 nm and 0.5 V gate voltage, measured at $V_\mathrm{DS}$ = 50 mV. Device area: 150 $\mu$m × 320 $\mu$m.}
\label{fig:Figure3}
\end{figure*}

Figure \ref{fig:Figure2}(a) presents the time-dependent source-drain current while modulating the laser ON and OFF in 60 s intervals at 785 nm. The shaded dark red regions indicate the laser ON periods. Measurements were performed at a fixed $V_\mathrm{DS}$ = 50 mV and $V_\mathrm{GS}$ ranging from -1 V to +1 V. At negative gate voltages, the current decreases when the light is ON, whereas at gate voltages equal to or above 0 V, the current increases under illumination. These results are consistent with the behavior observed in the transfer curves under illumination (Figure \ref{fig:Figure1}(c)), further confirming the gate-dependent nature of the photoresponse. The corresponding ${I_\mathrm{ph}}$ and $R$ values, extracted from these time-dependent curves as a function of gate voltage, are presented in the Supplementary Figure \ref{fig:FigureS9}. As expected, the maximum $R$ is reached at 0.5 V, in agreement with the trend reported in Figure \ref{fig:Figure1}(d). 

In addition to the gate-dependent behavior, the time evolution of the current during laser ON/OFF cycles reveals further details about the photoresponse dynamics. Across the full gate voltage range, when the laser is turned ON, the current exhibits a sharp initial increase followed by a slower rise until reaching a steady state. This behavior is indicative of two distinct photoresponse mechanisms \citep{ChengSciRep2013}: the initial rise can be attributed to the photogeneration and transfer of carriers (electrons or holes from CDs to SLG), assisted by the high mobility of graphene ($\mu \sim 7000$-$12000$ $\mathrm{cm^2\,V^{-1}\,s^{-1}}$; Supplementary Figure \ref{fig:FigureS10}). The slower rise that follows likely originates from charge trapping/detrapping processes in localized states \citep{AhnNanoAdv2021, SunAdvMater2012, KonstantatosNatNanotechnol2012}, either within the CDs or at the CD/SLG interface. These trapped charges accumulate over time and induce an intrinsic photogating effect \citep{ShinMicromachines2018}, gradually modulating the channel's conductivity. When illumination is switched OFF, the current drops, corresponding to the recombination of free carriers, and then decays more slowly due to the relaxation of trapped charges \citep{ChengSciRep2013}. 

\begin{figure*}[htbp!]
\centerline{\includegraphics[width=180mm]{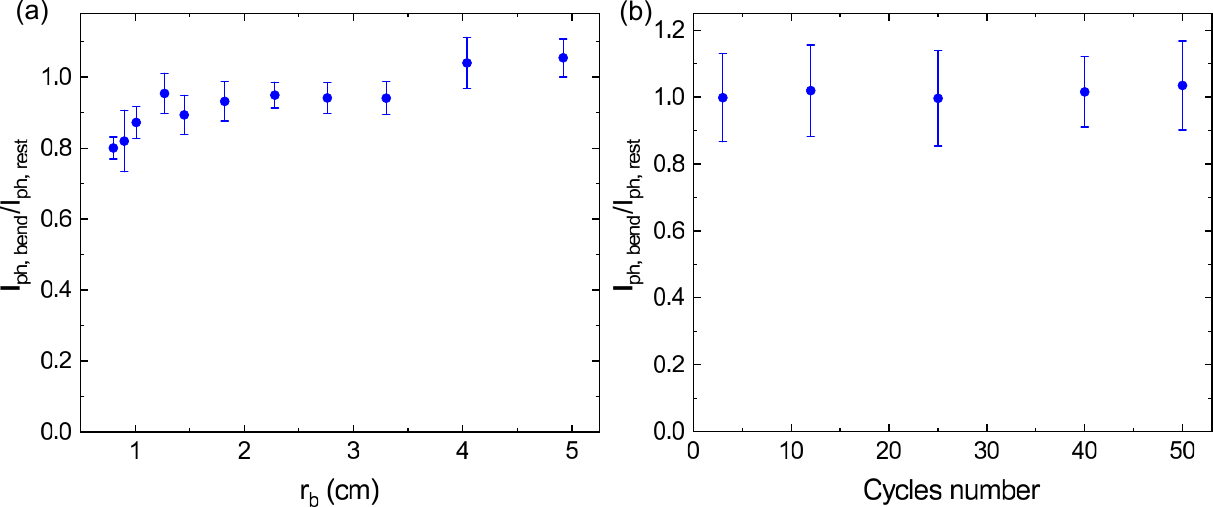}}
\caption{Normalized photocurrent response under (a) different bending radii and (b) repeated mechanical cycling.}
\label{fig:Figure4}
\end{figure*}

Measurements of the photoresponse versus time were performed using a digital oscilloscope and a waveform function generator. The function generator modulated the light source with a square-wave signal at a frequency of 100 mHz, enabling periodic ON/OFF illumination cycles. A constant $I_\mathrm{DS} = 50\ \mu$A was applied using a source-meter unit (SMU) operating in current-bias mode, while the resulting photovoltage ($V_\mathrm{ph}$) was recorded by the oscilloscope through a preamplifier with a gain of 10. This measurement configuration differs from that used in Figures \ref{fig:Figure1}(c) and \ref{fig:Figure2}(a), where a fixed $V_\mathrm{DS}$ was applied and $I_\mathrm{DS}$ was measured. The current-bias mode was chosen to directly monitor the dynamic voltage response of the device under modulated illumination. Figure \ref{fig:Figure2}(b) presents $V_\mathrm{ph}$ signals at various illumination wavelengths, under zero gate bias. The shaded magenta regions indicate the laser ON periods. For 785 and 642 nm illumination, the $V_\mathrm{ph}$ decreases upon illumination and recovers during the OFF state. In contrast, under 406 nm illumination, the $V_\mathrm{ph}$ increases with light exposure, indicating an opposite photoresponse polarity. These results are consistent with the transfer characteristics discussed earlier in Figure \ref{fig:Figure1}(c).

To further characterize the dynamic behavior of the device, the rise time ($\tau_\text{r}$) and the fall time ($\tau_\text{f}$) of the $V_\mathrm{ph}$ response were extracted using the standard 10\%–90\% method \citep{LongSciAdv2017}. Specifically, $\tau_\text{r}$ is defined as the time required for the signal to increase from 10\% to 90\% of its peak value, while $\tau_\text{f}$ corresponds to the time it takes for the signal to decrease from 90\% to 10\%. For each wavelength, the time interval between the points at which the signal reached 10\% and 90\% of its full amplitude (upon laser ON for rise time, and OFF for fall time) was measured. As shown in the Supplementary Figure \ref{fig:FigureS11} (a-c), $\tau_\text{r}$ and $\tau_\text{f}$ vary significantly with the excitation wavelength. The device exhibits the fastest rise time under 406 nm illumination ($\tau_\text{r}$ = 0.29 s), while slower responses are observed under 642 nm and 785 nm excitation ($\tau_\text{r}$ = 1.12 and 0.70 s, respectively). The fall times are comparatively longer across all wavelengths (2.06 s, 1.75 s, and 1.76 s for 406 nm, 642 nm, and 785 nm, respectively). These response times are suitable for wearable biometric sensors and environmental monitoring applications \citep{LapsaApplSci2024, TranNanotechnology2017, ChenChinesePhysB2018}.

Figure \ref{fig:Figure3}(a) displays the extracted photocurrent, $I_\mathrm{ph}$, and the corresponding $R$ as a function of source-drain voltage under 785 nm illumination for another device. The measurements were carried out at a fixed effective optical power of $\sim$0.2 mW. For each bias point, we recorded the time-dependent source-drain current while modulating the laser in 60 s ON/OFF intervals for two complete cycles. To extract $I_\mathrm{ph}$, we applied a 10\%–90\% method: current values were selected from the stable plateau regions within each light-ON and light-OFF interval, and their difference was taken to obtain the $I_\mathrm{ph}$ for each cycle. The final $I_\mathrm{ph}$ was calculated as the average of the values of the two cycles. The associated error bars were estimated as half the absolute difference between the two $I_\mathrm{ph}$ values, i.e., $\sigma_{I_\mathrm{ph}} = |I_{\text{ph,c1}} - I_{\text{ph,c2}}|/{2}$. As shown, both $I_\mathrm{ph}$ and $R$ increase monotonically with $V_\mathrm{DS}$, rising from approximately 3 $\mu$A and 13 mA/W at 50 mV to 14 $\mu$A and 64 mA/W at 250 mV. This enhancement is attributed to the increased electric field across the graphene channel at higher $V_\mathrm{DS}$, which accelerates the extraction of photocarriers, thereby reducing their transit time and the recombination probability \citep{LiuNatCommun2024}. Such behavior is characteristic of photoconductive detectors, where a stronger electric field enhances carrier collection efficiency \citep{LiuNatCommun2024}. 

Figure \ref{fig:Figure3}(b) presents the ($P_\mathrm{eff}$-dependent) $I_\mathrm{ph}$ under 785 nm illumination, measured at a fixed $V_\mathrm{DS}$ = 50 mV and $V_\mathrm{GS}$ = 0.5 V. $I_\mathrm{ph}$ values and related error bars were extracted with the same method as those of Fig. \ref{fig:Figure3}(a). A clear $I_\mathrm{ph}$ was detected for $P_\mathrm{eff}$ > 9 $\mu$W, where $I_\mathrm{ph}$ follows a power-law dependence, $I_\mathrm{ph} \propto P^{\alpha}$, with $\alpha \sim 1.20$, indicating a quasi-linear behavior \citep{MengAdvSci2023, HanNanoscaleAdv2021}.

\subsection{\label{Flex}Mechanical Flexibility Assessment}

To evaluate the mechanical robustness of the device and its optoelectronic stability under flexural stress, bending tests were performed using a vertical three-point bending stage (Deben Microtest, 300 N). During measurements, electrical connections to the device pads were established using silver paint and RoHS-compliant wires. This setup induces controlled deformation by applying force at three contact points, creating a circular arc with a well-defined bending radius $r_\mathrm{b}$. The radius was calculated using the geometric relation \citep{DeFazioACSNano2016}:
\begin{equation}
    r_\mathrm{b} = \frac{h^2 + \left(\frac{L}{2}\right)^2}{2h}
\end{equation}
where $L$ is the chord length between the two fixed supports and $h$ is the midpoint deflection (the vertical displacement of the arc).

During bending, the device was continuously illuminated, and the photocurrent $I_\mathrm{ph}$ was monitored in real time to assess any strain-induced degradation. The photocurrent recorded under bending $I_\mathrm{ph,bend}$ was normalized to the flat-state value $I_\mathrm{ph,rest}$ to evaluate the retention of performance. As shown in Figure \ref{fig:Figure4}(a), the device maintains stable optoelectronic functionality even under severe deformation, operating reliably down to the smallest tested bending radius of 0.8 cm without mechanical failure. The measurements show that the normalized $I_\mathrm{ph}$ remains within $\sim$15\% deviation for bending radii down to $\sim$1 cm, highlighting the excellent flexibility of the device. At the most extreme bending ($r_\mathrm{b}$ = 0.8 cm), the $I_\mathrm{ph}$ exhibits a slightly larger but still reversible reduction of $\sim$19.9\% $\pm$ 3.5\% compared to the flat state.

To further assess mechanical durability, cyclic bending tests were iteratively performed. The device was first characterized in the flat state, followed by repeated bending to a radius of approximately 1.3 cm for 50 consecutive cycles. As shown in Figure \ref{fig:Figure4}(b), the normalized $I_\mathrm{ph}$ remained highly stable throughout the cycling process. Across all bending cycles, deviations from the flat-state $I_\mathrm{ph}$ stayed within $\sim$3\%, confirming excellent mechanical resilience and demonstrating the device's reliability for flexible optoelectronic applications.

\subsection{\label{Skin}Skin-Compatibility Evaluation}

\begin{figure}[htbp!]
\centerline{\includegraphics[width=90mm]{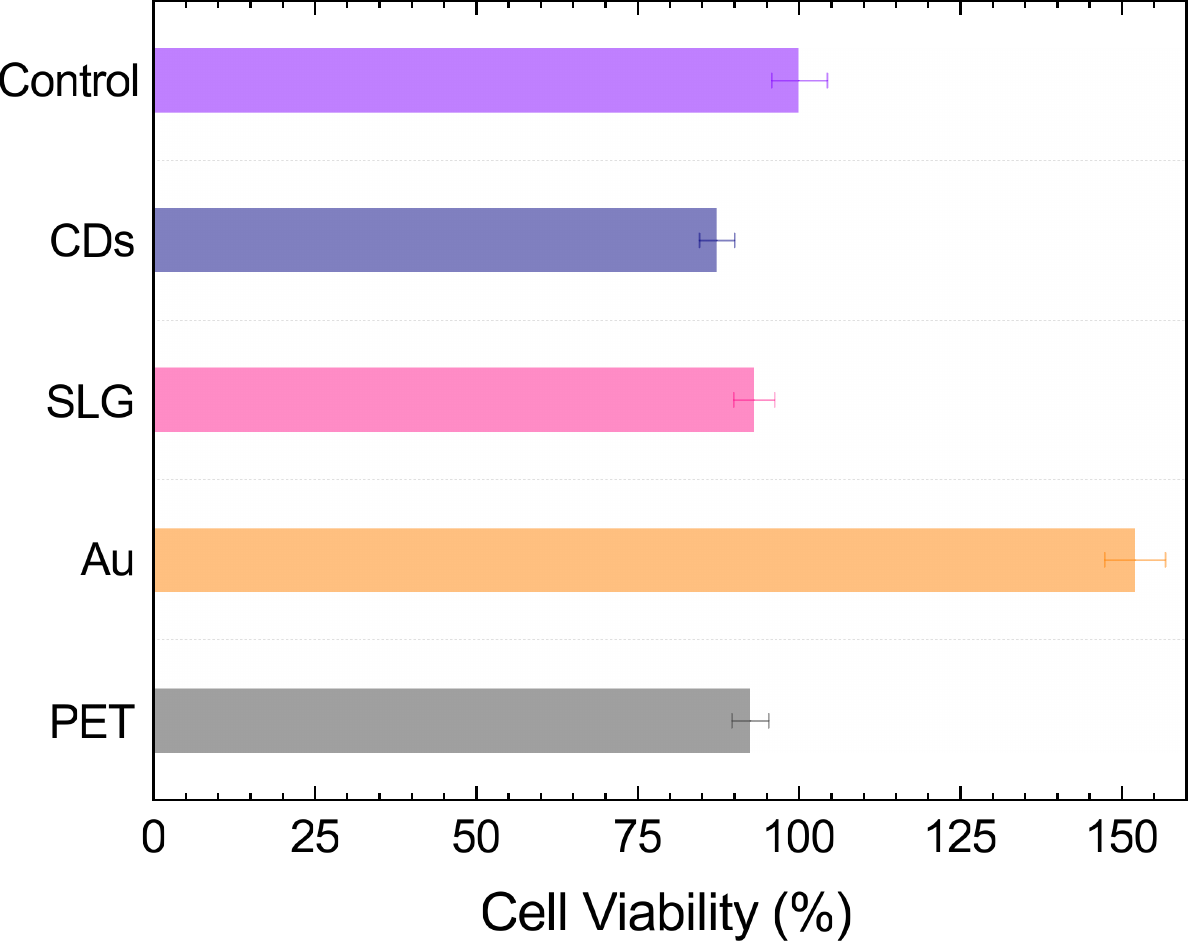}}
\caption{Cell viability of the CDs/SLG/PET device gated with CS-GL electrolyte and individual components evaluated for skin compatibility.}
\label{fig:Figure5}
\end{figure}

Skin irritation assays were performed using the SkinEthic\textsuperscript{\textnormal{TM}} RHE/S/17 model \citep{Episkin}, a reconstructed human epidermis generated from normal human keratinocytes cultured at the air-liquid interface on an inert polycarbonate membrane. This model closely replicates the histological architecture and barrier properties of native human epidermis, providing a standardized, reproducible, and reliable in vitro platform for toxicological assessments. The irritation test protocol was conducted according to the validated Episkin\textsuperscript{\textnormal{TM}} reference method \citep{VergaraJBiomed2023} with only minor modifications. A detailed description of the followed procedure is provided in the Experimental Section. Cell viability was assessed using the CellTiter-Glo® 3D Cell Viability Assay. The viability of treated tissues was normalized to that of the untreated control, defined as the 100\% viability reference. According to the manufacturer’s guidelines \citep{Episkin}, a viability above 50\% is considered indicative of non-irritant behavior. 

Across all tested samples, cell viability remained well above the 50\% threshold for cytotoxicity, with values ranging from 87.26\% (CDs) to 152.02\% (Au), as shown in Figure \ref{fig:Figure5}. These results highlight the preservation of epidermal cell function and confirm the overall skin-compatibility of the developed systems. The observed increase in cell viability beyond 100\% in the Au-treated sample may reflect a proliferative effect induced by gold released into the culture medium during the cytotoxicity assay. S. Lu et al. \citep{LuColloidsSurfB2010} demonstrated that low concentrations of gold nanoparticles ($\sim$5 ppm) significantly promoted keratinocyte proliferation in vitro, suggesting that similar mechanisms could underlie the enhanced viability observed in our study. Moreover, Lee et al. \citep{LeeBiotechnolBioprocessEng2022}  reported that human neonatal dermal fibroblasts (HDFn) exposed to gold nanorods at ultra-low (picomolar) concentrations exhibited a modest but significant increase in cell viability, likely driven by mild reactive oxygen species (ROS) generation that activates pro-growth signaling pathways.

\section{\label{End}Conclusions}

In conclusion, we have developed a broadband, flexible, and skin-compatible PD by integrating hydrothermally synthesized CDs with CVD-grown SLG on a PET substrate, gated by a biopolymer-based CS-GL electrolyte. In contrast to many existing flexible PDs that address only one or two key performance metrics, our device combines wide spectral response (UV–VIS–NIR), low-voltage operation (with peak performance at a gate bias of $\sim$0.5 V), mechanical robustness (stable performance at 0.8 cm bending radius), and verified skin compatibility. Three main innovations distinguish this work from the current state-of-the-art. First, we report the use of broadband-absorbing CDs, engineered through synthesis optimization to extend absorption well into the VIS-NIR region, without relying on toxic heavy metals, a significant step beyond conventional QDs or UV-limited CDs. Second, we introduce a CS-GL biopolymer electrolyte as a low-voltage, skin-safe gating medium, offering a scalable and biocompatible alternative to conventional ionic gels or solid-state dielectrics. Finally, the integration of these materials into a device architecture yields a device with responsivities of 0.19, 0.32, and 0.18 A/W at 406, 642, and 785 nm, respectively, competitive with other flexible PDs in this spectral range, but without compromising safety, cost, or mechanical performance. This work thus establishes a practical and scalable route to wearable PDs that satisfy the multiple, and often conflicting, requirements of real-world on-skin applications. Future work can further enhance performance, such as improving responsivity or extending spectral coverage, but this demonstration already marks a significant step toward next-generation on-skin PDs. 

\section{\label{Sec:Exp}Experimental Section}

\textit{Materials}: Citric acid (Merck Life Science S.r.l., Milano, Italy), urea (reagent grade, 98\%), chitosan (high molecular weight, Sigma-Aldrich), glycerol (ReagentPlus®, $\geq$99.0\%, Sigma-Aldrich), acetic acid (ACS reagent, $\geq$99.8\%, Sigma-Aldrich), and ammonium persulfate (ACS reagent, $\geq$98.0\%, Sigma-Aldrich). Polyethylene terephthalate (PET) films (Melinex® ST 507, heat stabilized, thickness: 125 $\mu$m) were purchased from König Folienzentrum (Germany). Poly(methyl methacrylate) (PMMA, AR‑P 672.045, E-Beam Resist PMMA 950K) was obtained from Allresist GmbH (Strausberg, Germany). SLG sheets (20 cm × 20 cm) grown on copper foil were purchased from Graphenea (San Sebastián, Spain). Deionized (DI) or ultrapure water was used in all experiments.

\textit{CDs Synthesis Procedure}: CDs were synthesized via a hydrothermal method using citric acid as the carbon source and urea as a nitrogen-rich dopant \citep{TaspikaRSCAdv2019, PermatasariACSApplNanoMater2018}. In a typical procedure, citric acid and urea were dissolved in 25 mL of ultrapure water and transferred to a Teflon-lined stainless-steel autoclave, then heated at 160 °C. After cooling to room temperature, the suspension was filtered and dialyzed against ultrapure water for 24 hours using a 1 kDa MWCO dialysis bag, with water changed every 12 hours. The final product was dried under reduced pressure. Three representative samples, shown in Figure \ref{fig:Figure1}(b), were prepared with varying precursor ratios and reaction times: CDs-1 (citric acid:urea = 1:0.5, 24 h), CDs-2 (1:0.5, 6 h), and CDs-3 (1:2, 8 h).

\textit{Transfer of CVD-Grown SLG onto PET substrate}: CVD-grown SLG on copper foil, purchased from Graphenea, was transferred onto PET substrates using a PMMA-assisted wet transfer method \citep{BaeNatNanotechnol2010}. First, a PMMA layer was spin-coated onto the graphene at 2000 rpm for 40 seconds and left to dry overnight at room temperature. The sample was then immersed in a 0.1 M ammonium persulfate solution for 5 hours to etch away the copper foil. Once the copper was fully dissolved, the resulting PMMA/graphene film was scooped from the solution using a glass slide and transferred into a DI water bath for 20 minutes. To ensure complete removal of residual etchant, the film was transferred to a second DI water bath for an additional 20 minutes. The PMMA/graphene film was then placed onto the PET substrate and dried on a hot plate at 60 °C. Finally, the PMMA layer was removed by immersing the sample in an acetone bath overnight, followed by rinsing with isopropyl alcohol (IPA) and final drying on a hot plate at 60 °C, completing the transfer process. 

\textit{Raman Characterization of SLG}: To investigate the quality of SLG, Raman spectra of SLG on Cu and SLG on Si/SiO$_2$ were obtained using a Renishaw inVia confocal Raman microscope (Renishaw, UK) with a 514 nm excitation laser (2.5 mW power), a grating of 1800 l/mm, and a 50x objective. Raman spectra of SLG on PET were collected using the same instrument equipped with a 785 nm excitation laser (5 mW power), a 1200 l/mm grating, and a 100x objective. 

\textit{CS-GL Electrolyte Preparation}: CS-GL Electrolyte was prepared by mixing 750 mg of chitosan with 25 mL of DI water, followed by the addition of 750 mg of glycerol. After thorough manual mixing to ensure uniform dispersion, 0.25 mL of acetic acid was added to facilitate chitosan dissolution. The resulting mixture was then subjected to ultrasonication at 45 °C to promote complete homogenization and dissolution of the components \citep{AhmadACSApplMater2025}.

\textit{Shadow Mask Electrode Patterning and Maskless Lithography for Graphene Devices}: Electrodes were fabricated on the SLG on PET, which had been previously transferred. Electrode patterning was achieved using a 150 $\mu$m‑thick aluminum shadow mask, purchased by Serigraph SNC, on which our electrode design was precision‑cut using a fiber laser. Physical vapor deposition (PVD) was performed using an electron-beam evaporator, where a 10 nm chromium (Cr) adhesion layer was first deposited, followed by a 50 nm gold (Au) layer to define the contact pads, followed by lift-off in acetone, rinse in isopropyl alcohol, and final drying with a nitrogen gun. To define the active graphene channel and remove excess graphene, we employed a maskless lithography system (Smart Print, model V2-TW5400, Microlight3D) equipped with a G-line UV source, and an Epson EH-TW5400 projector was used for direct photopatterning. A $\sim$2 $\mu$m thick layer of positive photoresist (S1818) was deposited to selectively protect the SLG only in the channel region bridging the source and drain electrodes. The patterned photoresist was then developed using an MF-351. The unprotected SLG area was etched using reactive ion etching (RIE), leaving behind only the desired graphene channel and removing shorts with the gates. Finally, the photoresist was removed with acetone, resulting in isolated SLG channels and lateral gate pads ready for subsequent electrical characterization. 

\textit{CDs Characterization}: The synthesized CDs were characterized using a variety of advanced spectroscopic and microscopic techniques. Raman spectra of the CDs in powder and film forms were obtained using a Renishaw inVia confocal Raman microscope (Renishaw, UK), using a 514 nm excitation laser (2.5 mW power), a 1800 l/mm grating, and a 50x objective. Fourier-transform infrared (FTIR) spectra were recorded with a Spectrum Two FTIR spectrometer (PerkinElmer, USA) to identify surface functional groups. UV–VIS absorption spectra were measured using a UV-2600 spectrophotometer (Shimadzu, Japan). Transmission electron microscopy (TEM) images were captured with a JEOL JEM-F200 microscope (cold FEG, JEOL, Japan) to examine morphology and estimate particle size.

\textit{Optoelectronic Measurements}: Electrical contacts were established via micromanipulators on the probe station (Model MPS150, FormFactor, USA). The devices were illuminated with continuous-wave lasers at wavelengths of 406 nm, 642 nm, and 785 nm, delivered by a multi-channel fiber-coupled laser source (Model MCLS1-CUSTOM, Thorlabs, USA). The output beams from the fiber channels were collimated using wavelength-matched fiber collimators (Models F810FC-405, F810FC-635, and F810FC-780; Thorlabs, USA) and then focused onto the active area of the device using a plano-convex N-BK7 lens (Model LA1708, focal length = 200 mm, Thorlabs, USA). Electrical characteristics were recorded using a precision source-measure unit (Model B2902A, Keysight Technologies, USA). Time-dependent photoresponse measurements were conducted using a source meter (Model 2400, Keithley Instruments, USA), a digital oscilloscope (Model RTM3004, Rohde \& Schwarz, Germany), and an arbitrary waveform generator (Model 33220A, Agilent Technologies, USA), with the signal recorded through a low-noise preamplifier (Model 1201, DL Instruments, USA) with a gain of 10. To assess the mechanical properties, the samples were subjected to bending tests using a bending stage (Deben, UK), which applied controlled forces to the devices. Impedance spectroscopy was carried out using a high-precision impedance/gain-phase analyzer (Model SI 1260, Solartron Analytical, UK), which measures the complex impedance over a wide frequency range (10 $\mu$Hz to 32 MHz). All measurements were conducted under ambient conditions.

\textit{Skin-Compatibility Testing Protocol}: On day 18 of RHE/S/17 culture, 300 $\mu$L of pre-warmed SkinEthic Maintenance Medium was added to each well of a sterile 24-well plate to allow for pre-incubation and tissue stabilization. Tissue inserts were gently removed from the transport agarose, cleaned to remove any residual gel, and transferred into the pre-filled wells, ensuring that no air bubbles were trapped beneath the membrane. The plate was then incubated for 4 hours at 37 °C under controlled conditions (5\% CO$_{2}$, 95\% humidity).
Following pre-incubation, the tissue inserts were transferred to a fresh 24-well plate containing 1 mL of pre-warmed SkinEthic Maintenance Medium. Test materials, in various physical forms, were then applied directly onto the tissue surface. The powdered sample consisted of CDs, while the solid samples included PET, SLG/PET, and Au. Each sample was gently placed on top of the epidermis. Three tissues were left untreated to serve as negative controls. All tissues, including controls, were incubated for 24 hours at 37 °C under standard atmospheric conditions (5\% CO$_{2}$, 95\% humidity).
After incubation, the test materials were removed, and tissues were thoroughly rinsed with phosphate-buffered saline (PBS) to eliminate any residual substances. The inserts were gently dried, transferred to a new 24-well plate, and each was filled with 300 $\mu$L of CellTiter-Glo® 3D Cell Viability Assay reagent, following the protocol by Barraza Vergara et al. \citep{VergaraJBiomed2023}. Luminescence was then measured using a Synergy H1 microplate reader. For each tissue, the luminescent solution was transferred to a 96-well plate, and measurements were performed using two technical replicates.
 
\section{\label{Ackn}Acknowledgements}
This work has been funded by the "Progetti di Ricerca di Ateneo SPIN - Supporting Principal Investigators - Avviso di selezione 2022" (CUP H75F22000030001) from Ca’ Foscari University of Venice, by the PRIN PNRR 2022 project "Continuous THERmal monitoring with wearable mid-InfraRed sensors (THERmIR)" (code P2022AHXE5, CUP 53D23007320001), by the INTERREG VI-A Italy–Croatia 2021–2027 project titled “Civil Protection Plan Digitalization through Internet of Things Decision Support System based Platform (DIGITAL PLAN)” (code ITHR020043, CUP H75E23000200005), and by the PRIMA 2023 project "Food value chain intelligence and integrative design for the development and implementation of innovative food packaging according to bioeconomic sustainability criteria (QuiPack)" (CUP H73C23001270005).

\section{\label{Dec}Declaration}

The authors declare no conﬂict of interest.

\clearpage
\widetext

\setcounter{equation}{0}
\setcounter{figure}{0}
\setcounter{table}{0}
\setcounter{page}{1}
\makeatletter
\renewcommand{\theequation}{S\arabic{equation}}
\renewcommand{\thefigure}{S\arabic{figure}}
\renewcommand{\thetable}{S\arabic{table}}
\renewcommand{\bibnumfmt}[1]{[S#1]}
\renewcommand{\citenumfont}[1]{S#1}
\newcounter{SIfig}
\renewcommand{\theSIfig}{S\arabic{SIfig}}

\begin{center}
\textbf{\large Supplementary Materials}
\end{center}

\section{\label{Table}Benchmark table}

\setlength{\arrayrulewidth}{0.2mm}

\begin{table}[htbp!]
\renewcommand{\arraystretch}{1.2}
\begin{tabularx}{\textwidth}{
  |>{\hsize=.15\hsize}X
  |>{\hsize=.15\hsize}X
  |>{\hsize=.15\hsize}X
  |>{\hsize=.15\hsize}X
  |>{\hsize=.15\hsize}X
  |>{\hsize=.17\hsize}X
  |>{\hsize=.10\hsize}X|
}
\hline
\textbf{Device} &
\textbf{Spectral Range} &
\textbf{Responsivity} &
\textbf{Response Time} &
\textbf{Bending} &
\textbf{Biocompatibility} &
\textbf{Reference} \\
\hline
CDs/SLG & 406-785 nm & 0.18-0.32 A/W & $\tau_\text{r}$ $\sim$ 0.29-1.12 s, $\tau_\text{f}$ $\sim$ 1.75-2.06 s & 8 mm radius, 50 cycles & Verified skin-compatible & This work \\
\hline
ZnO-based PD & 365 nm & $1.1 \times 10^{6}$ A/W & $\tau_\text{r}$ $\sim$ 10.1 s, $\tau_\text{f}$ $\sim$ 12.2 s & 40 mm radius & - & \citep{KaragiorgisNpjFlexElectron2025_s} \\
\hline
OPD & 400-950 nm & 0.47 A/W at 800 nm & $\tau_\text{r}$ $\sim$ 81 $\mu$s, $\tau_\text{f}$ $\sim$ 77 $\mu$s & 2 mm radius, 4000 cycles & Plant-compatible & \citep{SchrickxAdvOptMater2024_s} \\
\hline
PFBPPD & 350-770 nm & 0.15-0.2 A/W at 610-650 nm & $\tau_\text{r}$ $\sim$ 3.99 s, $\tau_\text{f}$ $\sim$ 39 ms & 5 mm radius, 50 cycles & Contains Pb (toxic) & \citep{AzamatACSApplOptMater2024_s} \\
\hline
QDPD & 1330 nm & - & 20 ns & 7 mm radius, 100 000 cycles & Contains Pb (toxic) & \citep{DengACSApplMaterInterfaces2025_s} \\
\hline
CsPbBr$_3$ QDs/Graphene & 520 nm & $\sim$82 000 A/W & $\tau_\text{r}$ $\sim$ 51 ms, $\tau_\text{f}$ $\sim$ 338 ms & 2.8 mm radius, 600 cycles & Contains Pb (toxic) & \citep{LiACSApplMaterInterfaces2025_s} \\
\hline
CQD/FF-Graphene & 200-400 nm & $4.66 \times 10^{6}$ A/W at 352 nm & - & 120 cycles & - & \citep{HsiaoJMaterChemC2024} \\
\hline
\end{tabularx}
\raggedright
\textbf{CDs – carbon dots; SLG – single-layer graphene; OPD – organic photodetector; PFBPPD – printed flexible bifacial perovskite photodetector; QDPD – colloidal quantum dot photodetector; CQD – carbon quantum dots; FF – fluorinated functionalization.}
\caption{Comparison of recently reported wearable photodetectors in terms of spectral range, responsivity, response time, mechanical flexibility, and biocompatibility.}
\label{tab:TableS1}
\end{table}

Table \ref{tab:TableS1} summarizes the key performance metrics of representative flexible photodetectors reported in recent literature. The devices are compared in terms of their spectral range, peak responsivity, time response (rise/fall time), bending stability (radius and/or cycles), and biocompatibility.

\section{\label{Raman}SLG Raman Analysis}

\begin{figure*}[htbp!]
\centerline{\includegraphics[width=180mm]{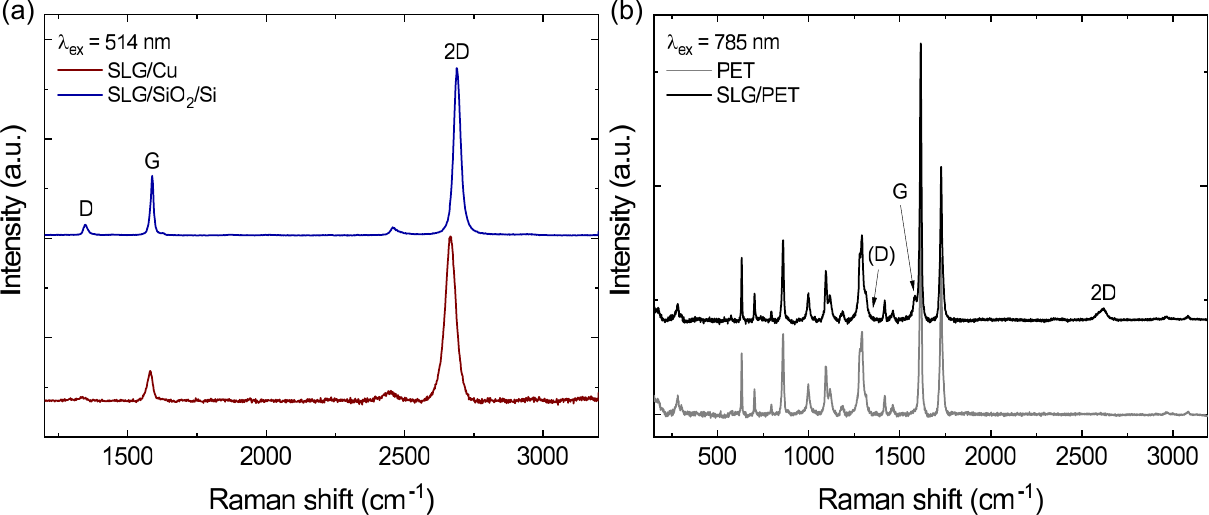}}
\caption{Raman spectra of (a) SLG/Cu and SLG/SiO$_2$/Si, and (b) PET and SLG/PET.}
\label{fig:FigureS1}
\end{figure*}

Raman spectroscopy was performed to investigate the vibrational signatures and structural quality of SLG before and after transfer. For SLG on Cu and after transfer onto a SiO$_2$/Si substrate (Figure \ref{fig:FigureS1}(a)), measurements were carried out using a 514 nm excitation laser with a power of 2.5 mW, a grating of 1800 l/mm, and a 50x objective. A fluorescence background was observed in the Raman spectrum of SLG/Cu, which was removed by baseline subtraction. Both spectra exhibit the characteristic features of graphene, namely the 2D, G, and D bands. The 2D peak (2665 cm$^{-1}$ for Cu and 2688 cm$^{-1}$ for SiO$_2$/Si) appears as a single Lorentzian, indicative of SLG\citep{FerrariNatNanotechnol2013}. The G band (1582 cm$^{-1}$ on Cu and 1588 cm$^{-1}$ on SiO$_2$/Si) corresponds to the E$_{2g}$ phonon at the Brillouin zone center \citep{FerrariNatNanotechnol2013}. The D band (1339 cm$^{-1}$ for Cu and 1347 cm$^{-1}$ for SiO$_2$/Si) arises from the breathing modes of sp$^2$ rings and requires a defect for its activation by double resonance \citep{FerrariNatNanotechnol2013, FerrariPhysRevLett2006}. The observed upshift of the bands after transfer is consistent with compressive strain induced during the transfer process \citep{LarsenMicroelectronEng2014}.  

For SLG on PET substrate (Figure \ref{fig:FigureS1}(b)), Raman spectra were collected using a 785 nm excitation laser with a power of 5 mW, a grating of 1200 l/mm, and a 100x objective. This selection followed preliminary measurements with a 514 nm laser, where fluorescence from the PET substrate adversely affected the quality of the SLG Raman signals, limiting the clarity of its bands. Using the 785 nm excitation mitigated this issue. A baseline subtraction was applied to further suppress the remaining background signal. The PET substrate exhibits broad polymer-related Raman bands \citep{BoerioJPolymSci1976}. Upon transferring SLG onto PET, the SLG-specific G band ($\sim$1583 cm$^{-1}$) and the 2D band ($\sim$2620 cm$^{-1}$) emerge distinctly above the polymer background, confirming the successful transfer of SLG onto PET. The observed redshift of the 2D band with 785 nm excitation, relative to 514 nm, is consistent with the known dispersion of the 2D mode in SLG arising from the double resonance Raman process \citep{WangChemPhysLett2013}.   

\section{\label{CharacCDs}Characterization of CD\MakeLowercase{s}}

\begin{figure*}[htbp!]
\centerline{\includegraphics[width=180mm]{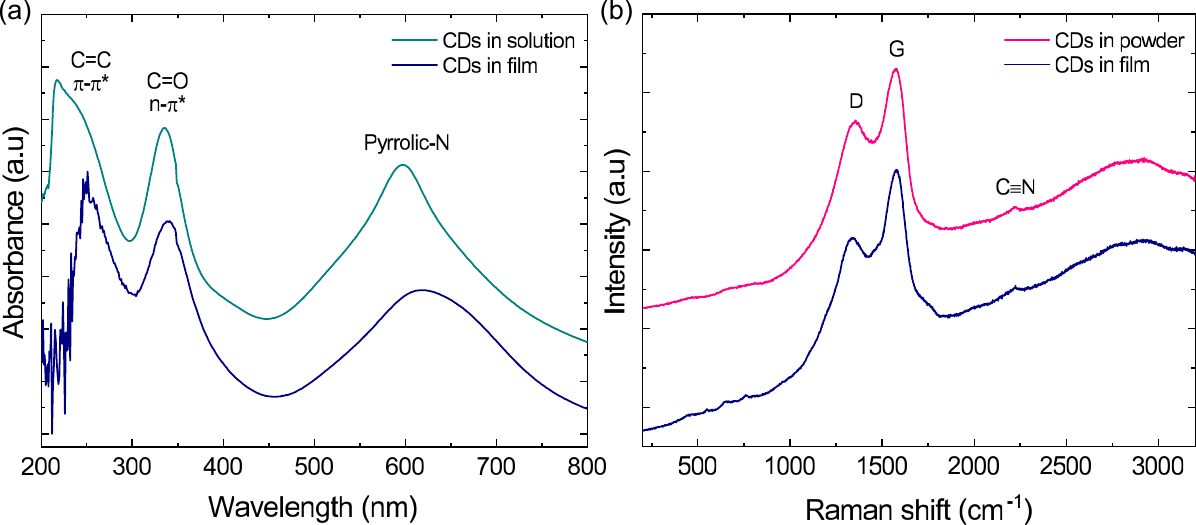}}
\caption{(a) UV-VIS absorption spectra, and (b) Raman spectra of CDs.}
\label{fig:FigureS2}
\end{figure*}

Figure \ref{fig:FigureS2}(a) illustrates the absorption spectra of the CDs in both aqueous dispersion and film forms. Upon film formation, the band at 600 nm becomes broadened and slightly red-shifted, reflecting the development of more delocalized $\pi$-conjugated systems across adjacent carbon domains, with these solid-state spectral modifications being consistent with $\pi$–$\pi$ stacking interactions, intermolecular aggregation, and the reduction of quantum confinement, all of which promote exciton delocalization and induce a bathochromic shift (red-shifting) \citep{DasMicromachines2021} in the absorption spectrum due to the narrowing of the effective bandgap in the aggregated phase.

Figure \ref{fig:FigureS2}(b) presents the Raman spectra of the synthesized CDs in both powder and film forms, acquired using a 514 nm excitation laser, a laser power of 2.5 mW, a grating of 1800 l/mm, and a 50x objective. The spectra prominently exhibit two principal peaks \citep{TanElecActa2015, YangJMaterChem2011, FerrariPhysRevB2001}: the D-band at approximately 1353 cm$^{-1}$, corresponding to the breathing modes of aromatic rings, activated by defects and disorder within the sp$^{2}$-hybridized carbon network. The intensity of the D-band is directly correlated with the defect density in the carbon structure. The second prominent feature, the G-band around 1575 cm$^{-1}$, originates from the E$_{2g}$ phonon at the Brillouin zone center and represents the in-plane stretching of C-C bonds in graphitic materials. This band is characteristic of sp$^{2}$-hybridized carbon atoms. Besides the D and G bands, both spectra also reveal a weaker feature around 2220 cm$^{-1}$, likely attributed to C$\equiv$N stretching vibrations from nitrile (–C$\equiv$N) groups \citep{MartonJspect2013}. 

\begin{figure*}[htbp!]
\centerline{\includegraphics[width=180mm]{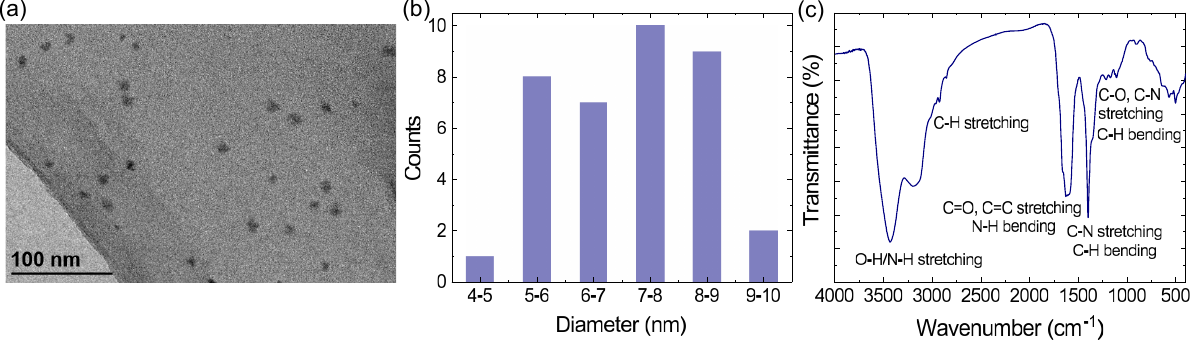}}
\caption{(a) TEM image, (b) size distribution, and (c) FTIR spectrum of CDs}
\label{fig:FigureS3}
\end{figure*}

Figure \ref{fig:FigureS3}(a) displays a TEM image of the synthesized CDs, revealing a uniform distribution of quasi-spherical nanoparticles. The absence of discernible lattice fringes further supports the amorphous nature of the carbon structure.  The corresponding histogram in Figure \ref{fig:FigureS3}(b) shows that the majority of particles fall within the 6-9 nm range. The average diameter was determined to be $7.11 \pm 1.33$ nm. 

Figure \ref{fig:FigureS3}(c) features the FTIR spectrum of the as-synthesized CDs. In the high-wavenumber region, a broad absorption band extending from approximately 3100 to 3500 cm$^{-1}$ is attributed to overlapping O-H and N-H stretching vibrations, indicating the presence of hydroxyl and amine functionalities introduced by citric acid and urea \citep{KhanSciRep2017}. These polar groups contribute significantly to the hydrophilicity and aqueous dispersibility of CDs. In addition, absorption bands observed between 2800 and 3000 cm$^{-1}$ can be assigned to symmetric and asymmetric C-H stretching vibrations of aliphatic -CH$_2$- and -CH$_3$ groups \citep{KhanSciRep2017} along with ammonium (RNH$_3^+$; R$_2$NH$_2^+$) groups \citep{CailottoACSApplMater2018}. The absorption region between 1500 and 1700 cm$^{-1}$ can be ascribed to the convolution of several vibrational transitions, including C=O stretching from carboxylic and amide groups, C=C stretching associated with sp$^{2}$ carbon domains, and N-H bending from amine or amide functionalities \citep{KurdekarMicrofluidNanofluid2016, ZhengJMaterSci2018}. The composite nature of this band suggests the coexistence of oxygen- and nitrogen-containing functional groups, as well as aromatic or conjugated carbon double bonds. The band observed near 1400 cm$^{-1}$ is likely due to C-N stretching and/or C-H bending vibrations, further confirming the incorporation of nitrogen into the surface structure \citep{ZhengJMaterSci2018}. The absorption features between 1000 and 1300 cm$^{-1}$ are attributed to C-O and C-N stretching vibrations, commonly associated with epoxy, alkoxy, and amide functionalities. Additionally, low-intensity bands below 900 cm$^{-1}$ are consistent with C-H out-of-plane bending \citep{KurdekarMicrofluidNanofluid2016}, supporting the presence of aromatic or heterocyclic motifs on the CD surface. Overall, the FTIR spectrum confirms the successful formation of oxygen- and nitrogen-doped CDs with chemically diverse surface functionalities.

\raggedbottom
\newpage

\section{\label{OptDevice}Optical Properties of Device Layers}

\begin{figure*}[htbp!]
\centerline{\includegraphics[width=90mm]{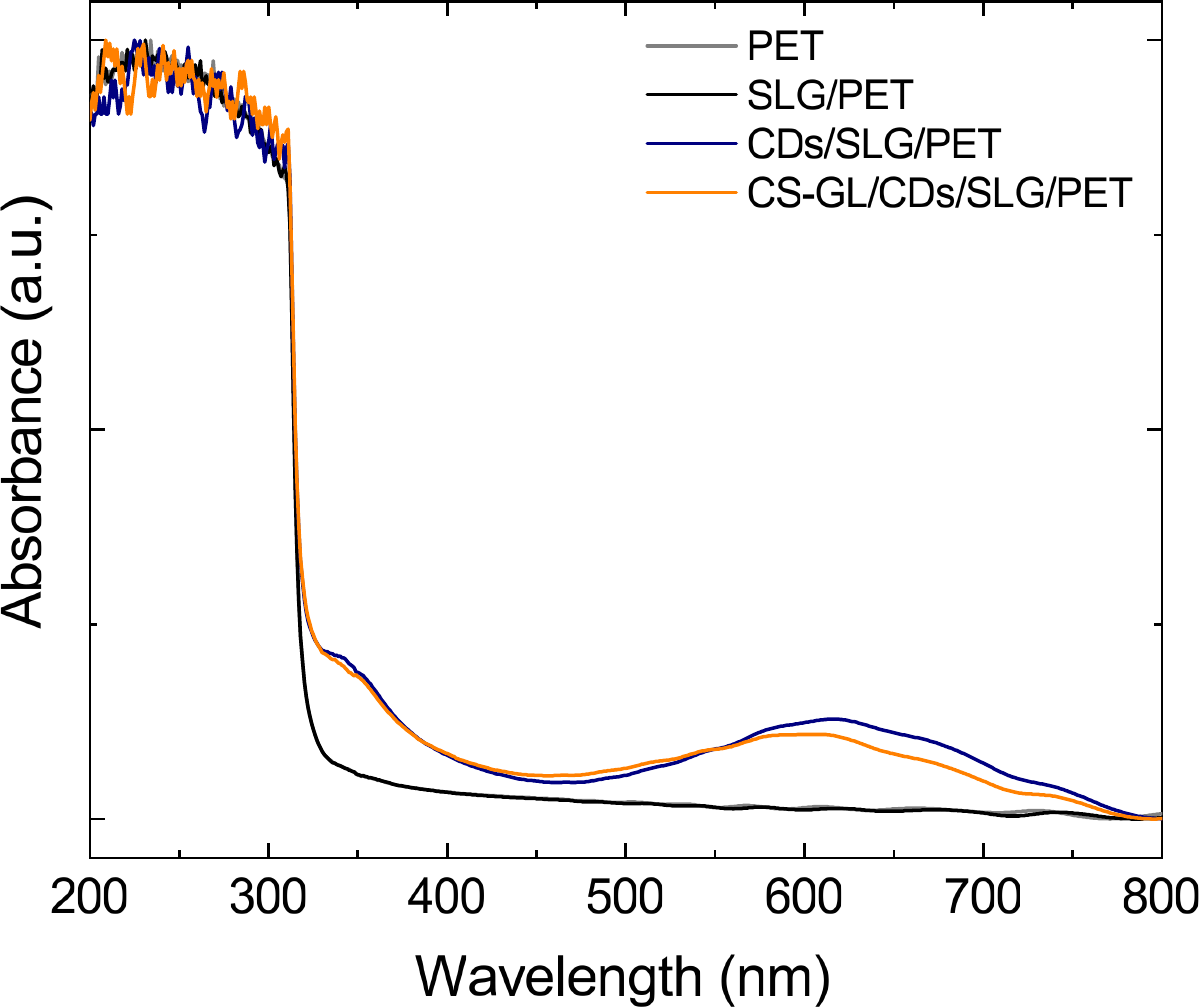}}
\caption{UV-VIS absorption spectra of device layers.}
\label{fig:FigureS4}
\end{figure*}

Figure \ref{fig:FigureS4} presents the UV-VIS absorption spectra of the device layers following sequential deposition. A clear enhancement in absorption is observed upon the introduction of CDs. In contrast, the SLG/PET sample shows minimal absorption, likely falling within the instrument's noise level and the strong optical background of the PET substrate.

\section{\label{SLGCDs}Complementary Roles of SLG and CD\MakeLowercase{s}}

\begin{figure*}[htbp!]
\centerline{\includegraphics[width=180mm]{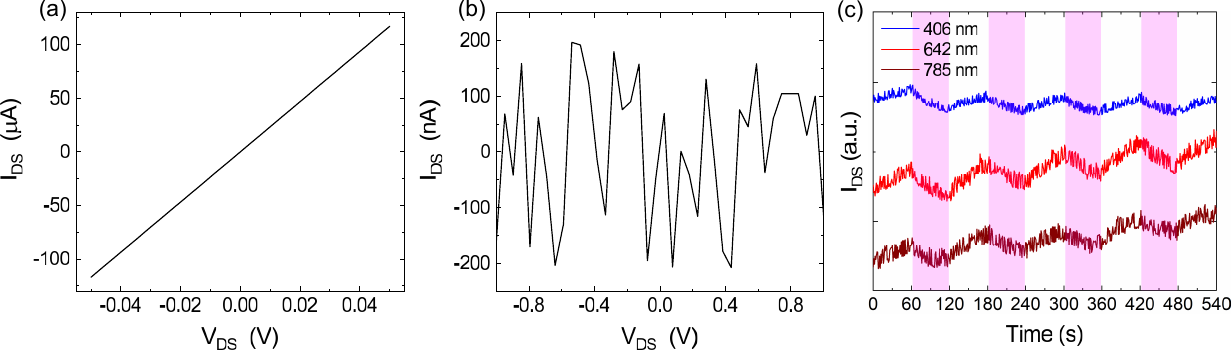}}
\caption{(a) Output characteristics of (a) SLG/PET and (b) CDs/PET under dark conditions. (c) Time-dependent photoresponse of SLG/PET under 406, 642, and 785 nm. Magenta regions correspond to laser ON states. Device area: 150 $\mu$m × 320 $\mu$m.}
\label{fig:FigureS5}
\end{figure*}

Figure \ref{fig:FigureS5}(a) shows the output characteristics of the SLG/PET device under dark conditions, confirming its linear I-V response, which is consistent with the expected ohmic contact behavior between SLG and the electrodes. Figure \ref{fig:FigureS5}(b) shows the corresponding I-V characteristics of the CDs/PET device, which exhibit negligible current under the applied bias, indicating that CDs alone are essentially non-conductive. 

To further examine the intrinsic photoresponse of SLG, we tested the SLG/PET device under periodic illumination with wavelengths of 406, 642, and 785 nm, modulated in 60 s ON/OFF cycles. The corresponding source-drain current dynamics, recorded at a constant drain bias of 50 mV and zero gate bias, are shown in Figure \ref{fig:FigureS5}(c). Only minor modulations in current were observed upon light exposure, reflecting the intrinsically low optical absorption of SLG.

\section{\label{Silver}Electrical Transfer Characteristics with Silver Electrodes}

\begin{figure*}[htbp!]
\centerline{\includegraphics[width=90mm]{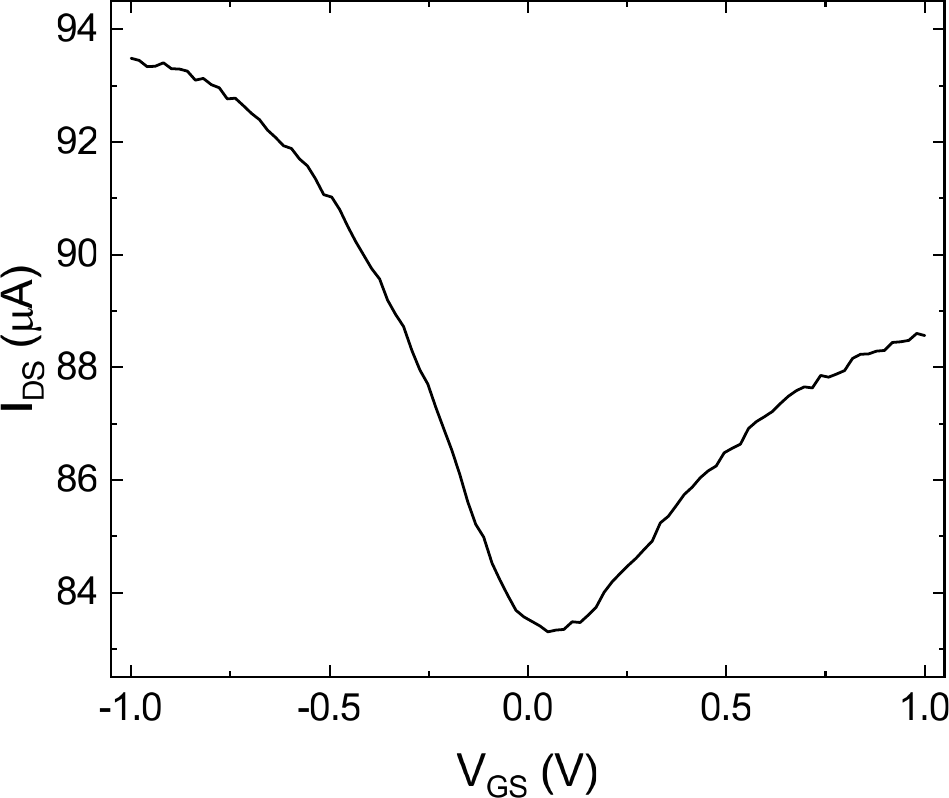}}
\caption{Transfer characteristics of a device measured in dark conditions with silver paint electrodes.}
\label{fig:FigureS6}
\end{figure*}

Figure \ref{fig:FigureS6} presents the transfer characteristics of the SLG/PET device measured under dark conditions using silver paint contacts. The curve exhibits the typical ambipolar behavior of graphene, with the Dirac point located near 0.05 V. These results serve as a baseline reference for comparison with devices fabricated using gold contacts.

\raggedbottom
\newpage

\section{\label{Noise}Power Spectral Density}

\begin{figure*}[htbp!]
\centerline{\includegraphics[width=90mm]{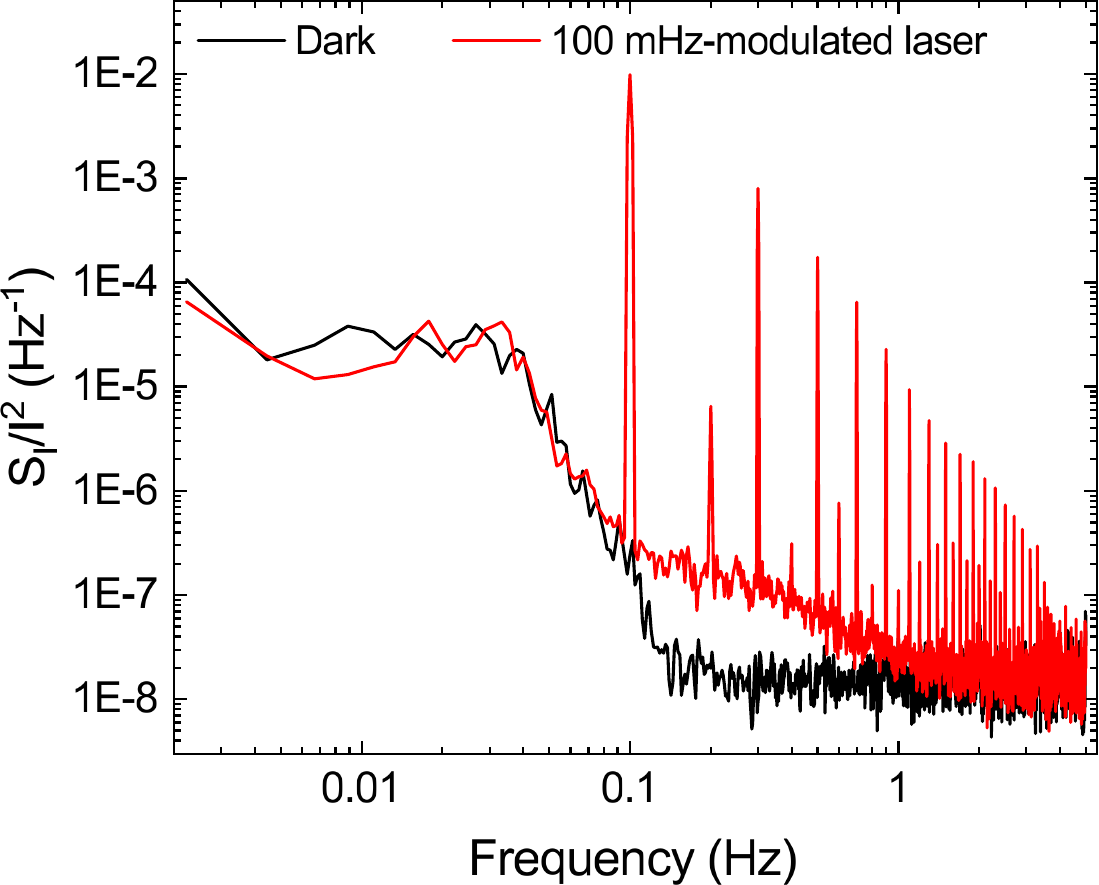}}
\caption{Normalized power spectral density (PSD) as a function of frequency from 5 mHz to 5 Hz. The black line shows the PSD under dark conditions. The red line corresponds to the PSD measured with a 100 mHz-modulated laser.}
\label{fig:FigureS7}
\end{figure*}

Figure \ref{fig:FigureS7} shows the normalized power spectral density (PSD) ($S_\text{I}/I^2$), as a function of frequency in the dark (black line) and under a 100 mHz-modulated laser illumination at 785 nm with an effective power $P_\mathrm{eff}$ of approximately 0.21 mW (red line). $S_\text{I}$ was calculated mathematically after biasing the device with a $V_\mathrm{DS}$ = 10 mV and recording $I_\mathrm{DS}$ for 1 hour at 0.1 s intervals, using Scipy library Welch's method \citep{WelchIEEETrans1967}. The calculated $S_\text{I}$ was then divided by the squared average of $I_\mathrm{DS}$ over the 1-hour measurement. At 100 mHz, with modulated light, $S_\text{I}/I^2$ is $\sim$ $9.90 \times 10^{-3}$ Hz$^{-1}$, over four orders of magnitude higher compared to the measurement in the dark, in which $S_\text{I}/I^2$ $\sim$ $2.41 \times 10^{-7}$ Hz$^{-1}$. It is worth noting that the current PSD exhibits distinct peaks not only at 100 mHz, corresponding to the modulation frequency, but also at its harmonics, highlighting the periodic nature of the photoresponse under illumination.

\raggedbottom
\newpage

\section{\label{EIS}Electrochemical Impedance}

\begin{figure*}[htbp!]
\centerline{\includegraphics[width=90mm]{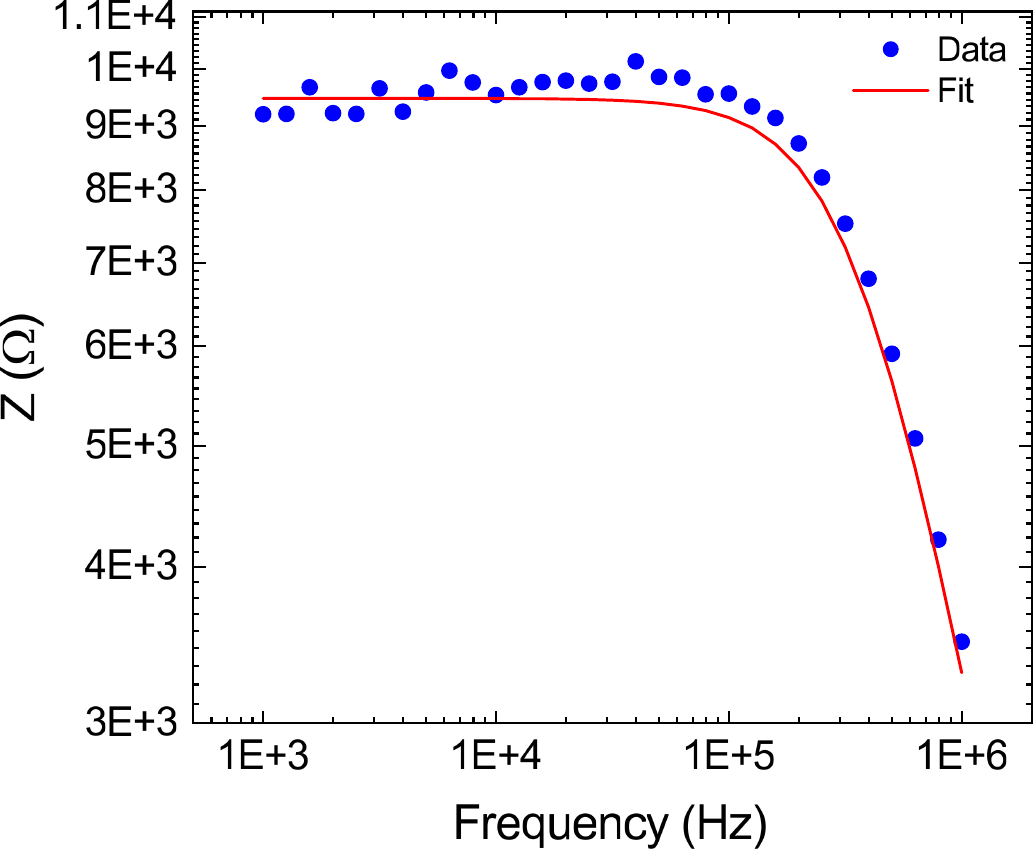}}
\caption{Frequency-dependent impedance of the device and equivalent circuit fit.}
\label{fig:FigureS8}
\end{figure*}

Figure \ref{fig:FigureS8} presents the frequency-dependent impedance of the device (blue dots) measured via EIS, along with the corresponding fit using an equivalent R-(RC) circuit model (red line). The good agreement between the measured data and the model confirms the suitability of the chosen circuit and supports the accurate extraction of the total capacitance of our device $C_\mathrm{tot}$.

\section{\label{PhR}Photocurrent and Responsivity Estimation}

\begin{figure*}[htbp!]
\centerline{\includegraphics[width=90mm]{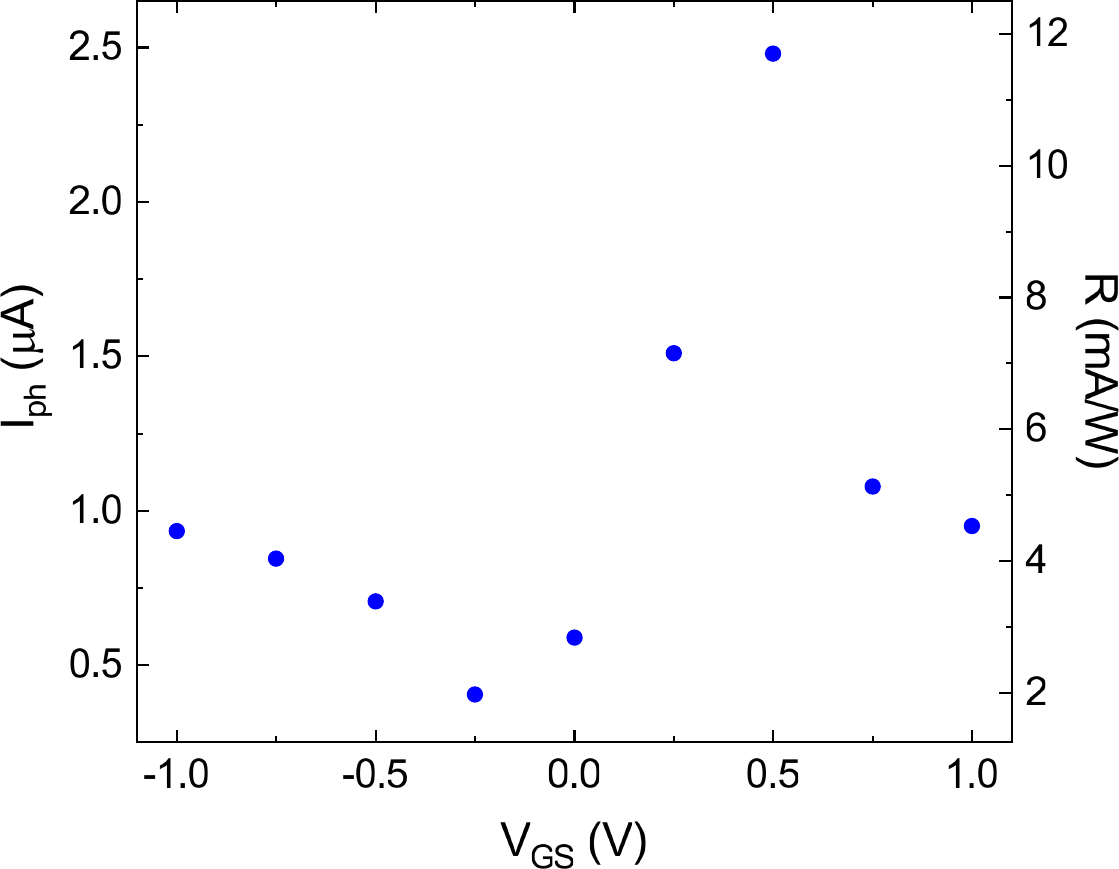}}
\caption{Gate-dependent photocurrent and responsivity estimated from Figure \ref{fig:Figure2}(a).}
\label{fig:FigureS9}
\end{figure*}

Figure \ref{fig:FigureS9} shows the gate-dependent $I_\mathrm{ph}$ (left axis) and the corresponding $R$ (right axis) extracted from the time-dependent measurements presented in Figure \ref{fig:Figure2}(a), performed under 785 nm laser illumination. The data confirm a clear gate-tunable photoresponse, with both $I_\mathrm{ph}$ and $R$ peaking at a gate voltage of 0.5 V.

\raggedbottom
\newpage

\section{\label{Time}Carriers Mobility}

\begin{figure*}[htbp!]
\centerline{\includegraphics[width=90mm]{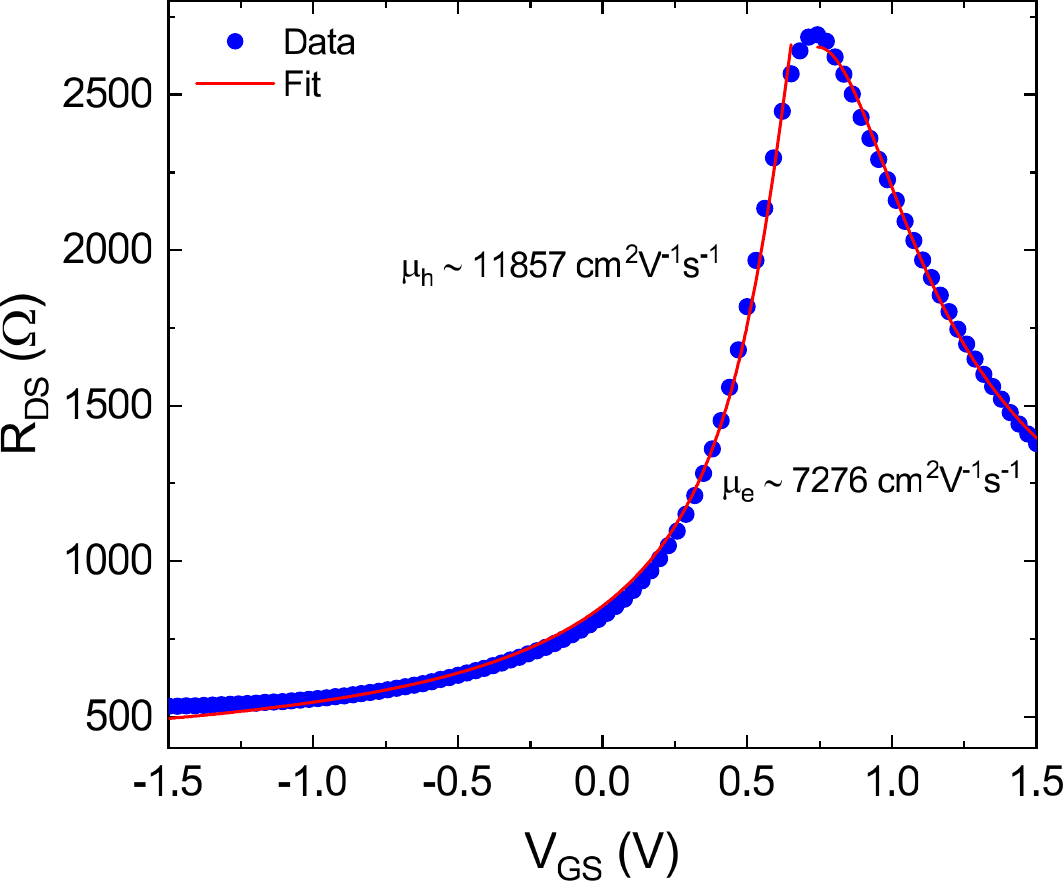}}
\caption{Gate voltage dependence of resistance with mobility extraction.}
\label{fig:FigureS10}
\end{figure*}

Figure \ref{fig:FigureS10} shows the resistance $R_\mathrm{DS}$, calculated from the transfer curve under dark conditions shown in Figure \ref{fig:Figure1}(c), measured at $V_\mathrm{DS} = 100\ \mathrm{mV}$, as a function of $V_\mathrm{GS}$. The experimental data (blue dots) are fitted (red curve) using a model that can estimate both hole ($\mu_\mathrm{h}$) and electron ($\mu_\mathrm{e}$) mobilities, using the following expression \citep{SmithACSNano2022}:
\begin{equation}
    R_\mathrm{DS} = R_0 + \left( \frac{L}{W} \right) \cdot \frac{1}{q \mu \sqrt{n_0^2 + n^2 }}
\end{equation}
where $R_0$ is the parasitic resistance, $L$ and $W$ are the channel length and width (150 $\mu$m and 320 $\mu$m, respectively), $q$ is the elementary charge, $\mu$ is the carrier mobility, $n_0$ is the residual carrier density, and $n$ is the gate-dependent carrier density, which is given by: $n = C_\text{tot} (V_{\text{GS}} - V_{\text{Dirac}})/{q}$, where $C_\text{tot}$ is the gate capacitance per unit area ($9.58 \times 10^{-8}~\mathrm{F{\cdot}cm^{-2}}$, measured via EIS), and $V_\mathrm{Dirac}$ is the Dirac point.

From the fitting procedure, the extracted mobilities are $\mu_\mathrm{h} \sim 11 857~\mathrm{cm^2V^{-1}s^{-1}}$ for holes and $\mu_\mathrm{e} \sim 7276~\mathrm{cm^2V^{-1}s^{-1}}$ for electrons. These values are comparable to those reported for CVD graphene on plastic substrates \citep{Khan2DMater2024, Hempel2DNanoscale2018}.

\raggedbottom
\newpage

\section{\label{Time}Time Response Calculation}

\begin{figure*}[htbp!]
\centerline{\includegraphics[width=180mm]{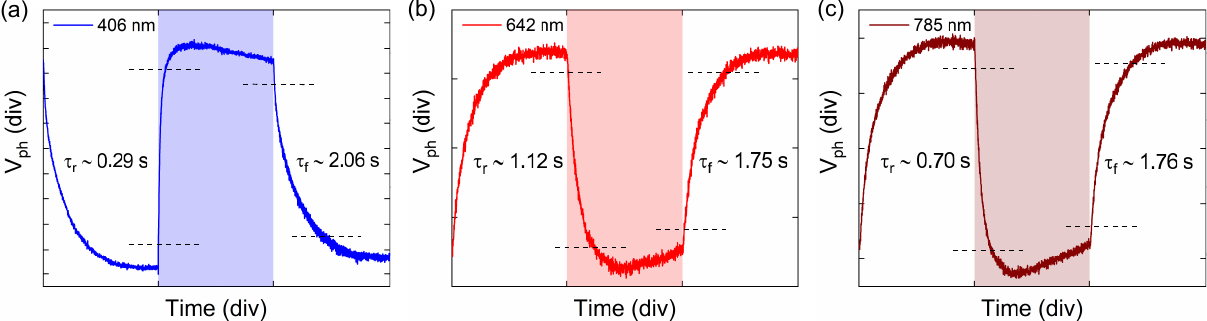}}
\caption{(a–c) Rise and fall time analysis of the photovoltage response under 100 mHz modulated illumination at 406 nm, 642 nm, and 785 nm, respectively, using the 10\%-90\% method. The horizontal scale is 5 s/div, and the vertical scale is 10 mV/div. Shaded regions correspond to laser ON states.}
\label{fig:FigureS11}
\end{figure*}

Figure \ref{fig:FigureS11} shows the rise and fall time analysis of the photovoltage response under 100 mHz modulated illumination at 406, 642, and 785 nm, extracted using the 10–90\% method. The rise times are 0.29 s, 1.12 s, and 0.70 s for 406, 642, and 785 nm, respectively, while the corresponding fall times are 2.06 s, 1.75 s, and 1.76 s.

\raggedbottom


\end{document}